\documentclass[11pt]{article}
\usepackage{graphicx} 
\usepackage[utf8]{inputenc}
\usepackage{titlesec}
\titleformat{\section}
{\normalfont\scshape}
{\thesection.}{0.5em}{\centering}
\titleformat{\subsection}
{\small\scshape}
{\thesubsection.}{0.5em}{\centering}
\titleformat{\subsubsection}
{\small\scshape}
{\thesubsubsection.}{0.5em}{\centering}
\usepackage{tikz-cd}
\usepackage{tikz}
\usepackage[left=3.5cm,right=3.5cm,top=3.5cm,bottom=3.5cm]{geometry}
\usepackage{comment}
\usepackage{amsfonts}
\usepackage{indentfirst}
\renewcommand{\abstract}{\small{\section*{\abstractname}}}
\usepackage{amsmath}
\usepackage{hyperref}
\hypersetup{
    colorlinks,
    citecolor=blue,
    filecolor=blue,
    linkcolor=blue,
    urlcolor=blue
}
\usepackage{amssymb}
\usepackage{amsthm}
\usepackage{enumerate}
\usepackage{mathrsfs}
\newtheorem{theorem}{Theorem}
\newtheorem{lemma}{Lemma}
\newtheorem{proposition}{Proposition}

\newtheorem*{corollary*}{Corollary}
\newtheorem*{theorem*}{Theorem}
\newtheorem{definition}{Definition}

\theoremstyle{remark}
\newtheorem{remark}{Remark}
\newtheorem{example}{Example}

\usepackage{stmaryrd}

\usepackage{enumitem}

\usepackage[
backend=biber,
style=numeric,
sorting=ynt
]{biblatex}
\title{\Large{\textbf{A General Theory of Risk Sharing}}}
\author{\normalsize{\textsc{Vasily Melnikov}}}
\date{\normalsize{\textsc{November 2026}}}

\begin{document}

\maketitle
\begin{abstract}
We introduce a new paradigm for risk sharing that generalizes earlier models based on discrete agents and extends them to allow for sharing risk within a continuum of agents. Agents are represented by points of a measure space and have potentially heterogeneous risk preferences modeled by risk measures on a separable probability space. We derive the dual representation of the value function using a Strassen-type theorem for the weak-star topology and provide a characterization of the acceptance set using Aumann integration. These results are illustrated by explicit formulas when risk preferences are within the family of entropic and expected shortfall risk measures, and applications to Pareto efficiency in large markets.
\end{abstract}
\section{Introduction}\label{sec:introduction}
A significant literature studies the risk sharing problem: can one distribute a risk among finitely many agents, such that the total risk is minimized? Mathematically, for a given loss $\mathcal{X}$, this corresponds to the optimization problem
\begin{equation}\label{eq:min-problem}
    \sum_{a\in A}\varrho_{a}(X_{a})\longrightarrow\mathrm{min!}
\end{equation}
subject to $\sum_{a\in A}X_{a}=\mathcal{X}$, where $(X_{a})_{a\in A}$ is a potential allocation of risk, modeled as a family of bounded random variables on a probability space $(\Omega,\mathscr{F},\mathbb{P})$. In (\ref{eq:min-problem}), the finite set $A$ represents the space of agents, and is the index set for a collection $(\varrho_{a})_{a\in A}$ of risk measures, with $\varrho_{a}$ defining the preferences of agent $a$.
\par
The risk sharing problem has a myriad of applications. In particular, optimal risk sharing plays an important role in the theoretical underpinnings of capital requirements and capital adequacy tests (see \cite{caprequirfs}), and has practical consequences for the Solvency II directive regulating European insurance firms (see \cite{solvencyiibad,solvencyiimf}). Furthermore, firms subject to regulations may seek to avoid requirements by dividing their assets, leading to a minimization problem of the same form as (\ref{eq:min-problem}) (see \cite{regsgetarbed}). In a more indirect manner, solutions to the risk sharing problem characterize Pareto efficiency (see \cite{ghossoubzhu2025}).
\par
However risk sharing is applied, the risk sharing framework implicit in (\ref{eq:min-problem}) is not flawless; since $A$ is finite, each agent has a non-negligible impact on the model. Though appropriate for ``too big to fail'' banks, such impacts ignore the diffuse nature of smaller financial institutions, whether it be community banks, credit unions, or individual investors. These actors may have diverse considerations and preferences most accurately modeled—at least, as an approximation—by continuum models, where any one actor has a negligible impact on the model. Continuum approximations have precedence in the economics and game theory literature (see, for instance, the seminal paper of Aumann \cite{aumanncontinuum}, or the theory of mean field games \cite{mfg}).
\par
While the above concerns justify including continuum agent models, they do not justify excluding all aspects of discrete agent spaces. In particular, the era of ``too big to fail'' has revealed the outsized influence of select financial institutions on the global market, even if sometimes counterbalanced by the combined impetus of smaller actors. Thus, it is necessary to consider both discrete and continuous agent spaces, potentially at the same time, reflecting the fact that some agents have essentially no impact on the market, while others may have disproportionate influence.
\par
We therefore adopt an arbitrary finite complete measure space $(A,\mathscr{A},\mu)$ as an agent space, where finiteness refers to the assumption that $0<\mu(A)<\infty$. The measure $\mu$ may be purely atomic (corresponding to a discrete agent space), non-atomic (corresponding to a continuum of agents), or a mix between the two—allowing one to model a wide range of circumstances.
\par
The cost of this universality is an increase in mathematical technicality. Some allocations of risk must be excluded for failing to satisfy measurability. Furthermore, since spaces of random variables are often infinite dimensional, there are multiple ways to choose a notion of measurability for allocations even for a fixed $\sigma$-algebra on $A$. Nor is integration, the replacement of the finite sums in (\ref{eq:min-problem}) and the associated constraint, easily assimilated into the theory; infinite dimensional spaces support multiple versions of the Lebesgue integral. We adjudicate each of these issues, establishing a unified mathematical framework to answer the problems of risk sharing when agents form a general measure space.
\par
Allowable allocations $(X_{a})_{a\in A}$ are assumed to be measurable, in a notion of measurability derived from the weak-star topology on $L^{\infty}(\mathbb{P})$, and integrable in the sense of Gelfand (see Definition \ref{def:gelf-int}). Risk preferences are represented by a collection of risk measures $(\varrho_{a})_{a\in A}$, which must also satisfy a measurability condition. More precisely, for every measurable allocation $(X_{a})_{a\in A}$, the mapping $a\longmapsto\varrho_{a}(X_{a})$ must be measurable. Once such assumptions are made, and an initial risk $\mathcal{X}\in L^{\infty}(\mathbb{P})$ is fixed, the minimization problem (\ref{eq:min-problem}) can be stated as
\begin{equation}\label{eq:new-min-problem}
    \int_{A}\varrho_{a}(X_{a})\mu(da)\longrightarrow\mathrm{min!}
\end{equation}
subject to the Gelfand integral of $(X_{a})_{a\in A}$ existing and equaling $\mathcal{X}$, potentially in addition to some other constraints.
\par
If (\ref{eq:new-min-problem}) is considered without constraints, we derive an explicit expression for the convex conjugate of the value function under general assumptions (see Theorem \ref{thm:dual-rep-represent} and Appendix \ref{sec:prf-of-thm-rep}). The formulas mimic the discrete case, replacing sums with integrals:
\begin{theorem*}
    Suppose that the minimization problem (\ref{eq:new-min-problem}) is considered without additional constraints and $\varrho(\mathcal{X})$ denotes the value function of the problem. If $(\varrho_{a})_{a\in A}$ are risk measures satisfying the Lebesgue property, $(\varrho_{a})_{a\in A}$ satisfies a measurability condition, $\int_{A}\vert{\varrho_{a}(0)}\vert\mu(da)<\infty$, and the value function is finite for all $\mathcal{X}$, then the value function is a risk measure, and the convex conjugate $\varrho^{\ast}$ satisfies
    \begin{equation*}
        \varrho^{\ast}(\mathbb{Q})=\int_{A}\varrho_{a}^{\ast}(\mathbb{Q})\mu(da)
    \end{equation*}
    for any probability measure $\mathbb{Q}\ll\mathbb{P}$, where $\varrho^{\ast}_{a}$ denotes the convex conjugate of $\varrho_{a}$ for each $a\in A$. In particular,
    \begin{equation*}
        \varrho(\mathcal{X})=\sup_{\mathbb{Q}\ll\mathbb{P},\mathbb{Q}(\Omega)=1}\left(\mathbb{E}^{\mathbb{Q}}(\mathcal{X})-\int_{A}\varrho_{a}^{\ast}(\mathbb{Q})\mu(da)\right).
    \end{equation*}
\end{theorem*}
Proving the above formulas requires establishing a Strassen-type theorem for the weak-star topology, which we do in Appendix \ref{sec:tech}.
\par
As an illustration of the general theory developed for solving (\ref{eq:new-min-problem}) in the unconstrained case, we give some concrete examples in \S\ref{sec:examples}, in particular for risk preferences within the family of entropic or expected shortfall risk measures. The formulas generalize previous results from the discrete case (including those of \cite{quantile-share} and \cite{righi-op}). These are derived as a special case of formulas for dilations and inflations of a fixed risk measure. The former family is known and subsumes entropic risk measures, while the latter is a new definition, and includes expected shortfall as a special case. Although the risk sharing problem is always well-posed for dilated risk measures (see Theorem \ref{thm:dilate-share}), we delineate sufficient conditions for the ill-posedness of the risk sharing problem for inflations of a fixed risk measure (see Theorem \ref{thm:attain-notattain} and Appendix \ref{sec:prf-attain-notattain}).
\par
In \S\ref{sec:pareto}, we show the problem (\ref{eq:new-min-problem}) is related to Pareto efficiency. More precisely, if the value function is finite, an allocation $(X_{a})_{a\in A}$ of $\mathcal{X}$ is Pareto efficient if, and only if, $(X_{a})_{a\in A}$ is a solution to the problem (\ref{eq:new-min-problem}). As a consequence, the ill-posedness results from \S\ref{sec:examples} imply economies where agents have preferences given by expected shortfall rarely have Pareto efficient allocations if there are a continuum of agents.
\section{Notation and Preliminaries}
Fix a complete measure space $(A,\mathscr{A},\mu)$, where $0<\mu(A)<\infty$. $A$ is the agent space, and elements $a\in A$ are agents. The spaces $L^{1}(\mu)$ and $L^{\infty}(\mu)$ carry their usual meaning.
\begin{example}\label{ex:finite}
    A typical model for agents in the risk sharing literature is $A=\{1,\dots,N\}$, where $N\in\mathbb{N}$ is the number of agents. Equipping $A$ with the $\sigma$-algebra $2^{\{1,\dots,N\}}$ and the counting measure assimilates this model into our framework.\qed
\end{example}
\begin{example}[Aumann \cite{aumanncontinuum}]\label{ex:aumman}
    For $A=[0,1]$, let $\mathscr{A}$ be the Lebesgue $\sigma$-algebra on $A$, and let $\mu$ be the normalized Lebesgue measure on $\mathscr{A}$. Such a choice of the triple $(A,\mathscr{A},\mu)$ represents a continuum of agents, each with negligible individual impact on the model.\qed
\end{example}
\begin{example}[Shapley \cite{shape}]\label{ex:shapely}
    For $A=[0,1]$, let $\mathscr{A}$ be the Lebesgue $\sigma$-algebra on $A$. Denoting by $\lambda$ the normalized Lebesgue measure on $[0,1]$, define $\mu=\lambda+\delta_{0}+\delta_{1}$, where $\delta_{i}$ is the Dirac measure centered at $i$ ($i=0,1$). This corresponds to an agent space with two large agents, and infinitely many small agents, such that the combined force of the smaller agents is equal to half the combined force of the larger agents.
    \qed
\end{example}
To the author's knowledge, neither of the agent spaces suggested by Example \ref{ex:aumman} or Example \ref{ex:shapely} has been considered by the risk measures literature before.
\par
Each agent faces uncertainty, which is modeled by a separable probability space $(\Omega,\mathscr{F},\mathbb{P})$. The state of the world is completely described by a corresponding point $\omega\in\Omega$. $\mathscr{F}$ represents the amalgamation of information communicated about the state of the world by various observables.
\par
Separability of $\mathbb{P}$ means that $\mathscr{F}$ equipped with the pseudometric $(B,C)\longmapsto\mathbb{P}(B\triangle C)$ is a separable pseudometric space. This technical assumption is crucial for our later arguments, especially those related to measurable selectors—for example, see Appendix \ref{sec:tech}, which assumes a separable ambient space. The assumption is usually not restrictive (being satisfied by any Lebesgue-Rokhlin space), and even when not satisfied, risk sharing can often be reduced to the separable case anyways (see \cite[Theorem 3.1(vi)]{law-inv-util-conv-dual}).
\par
The spaces $L^{1}(\mathbb{P})$ and $L^{\infty}(\mathbb{P})$ carry their usual meaning as spaces of contingent payoffs, although we adopt the convention that $\mathcal{X}\geq0$ represents a loss of magnitude $\mathcal{X}$. By the aforementioned separability assumption, $L^{1}(\mathbb{P})$ is separable. $\mathscr{M}_{\mathbb{P}}$ will denote the set of absolutely continuous probability measures $\mathbb{Q}\ll\mathbb{P}$.
\subsection{Allocations}\label{subsec:alloc}
A function $(X_{a})_{a\in A}$ from $A$ to $L^{\infty}(\mathbb{P})$, assigning to each agent $a$ some contingent payoff $X_{a}\in L^{\infty}(\mathbb{P})$, is called an allocation. Applying an integration theory to such functions requires making measurability assumptions. To this end, let us introduce a notion of measurability.
\begin{definition}
    An allocation $(X_{a})_{a\in A}$ is said to be $\mathscr{A}$-measurable if, for each $\mathcal{Y}\in L^{1}(\mathbb{P})$, the function $a\longmapsto\mathbb{E}^{\mathbb{P}}(X_{a}\mathcal{Y})$ is $\mathscr{A}$-measurable.
\end{definition}
Equipped with the above notion, we may define an integration theory for allocations.
\begin{definition}\label{def:gelf-int}
    An $\mathscr{A}$-measurable allocation $(X_{a})_{a\in A}$ is said to be Gelfand integrable if, for each $\mathcal{Y}\in L^{1}(\mathbb{P})$, the $\mathscr{A}$-measurable function $a\longmapsto\mathbb{E}^{\mathbb{P}}(X_{a}\mathcal{Y})$ is $\mu$-integrable.
\end{definition}
If $(X_{a})_{a\in A}$ is Gelfand integrable, for each $B\in\mathscr{A}$, there exists a unique element $\mathcal{Z}_{B}\in L^{\infty}(\mathbb{P})$ such that
\begin{equation*}
        \mathbb{E}^{\mathbb{P}}\left(\mathcal{Z}_{B}\mathcal{Y}\right)=\int_{B}\mathbb{E}^{\mathbb{P}}\left(X_{a}\mathcal{Y}\right)\mu(da),
\end{equation*}
for each $\mathcal{Y}\in L^{1}(\mathbb{P})$ (see \cite[p. 430]{aliprantis-inf-dim}). $\mathcal{Z}_{B}$ is called the Gelfand integral of $(X_{a})_{a\in A}$ over $B$, and is denoted $\int_{B}X_{a}\mu(da)$. By construction, Gelfand integrals are always elements of $L^{\infty}(\mathbb{P})$.
\par
Subsequently, we consider the problem of distributing a risk $\mathcal{X}\in L^{\infty}(\mathbb{P})$ among the agent space $A$. The above integration theory allows us to formalize the allowable distributions.
\begin{definition}
    An $\mathscr{A}$-measurable allocation $(X_{a})_{a\in A}$ is said to be $\mathcal{X}$-feasible if $(X_{a})_{a\in A}$ is Gelfand integrable, and $\mathcal{X}=\int_{A}X_{a}\mu(da)$. The set of $\mathcal{X}$-feasible allocations is denoted $\mathbb{A}(\mathcal{X})$.
\end{definition}
Formulating feasibility via an integral constraint is standard when agents are modeled as a measure space (see \cite[Definition 5.2(i)]{sagara-variational}). If $A=\{1,\dots,N\}$, all singletons are $\mathscr{A}$-measurable, and $1=\mu(\{1\})=\dots=\mu(\{N\})$ as in Example \ref{ex:finite}, $\mathcal{X}$-feasibility reduces to $\mathcal{X}=\sum_{a\in A}X_{a}$.
\subsection{Risk Preferences}
Each agent has risk preferences, which are modeled by a risk measure.
\begin{definition}\label{def:risk-measure}
    A risk measure is a functional $\varrho:L^{\infty}(\mathbb{P})\longrightarrow\mathbb{R}$ satisfying properties (\ref{it:risk-monotone}), (\ref{it:risk-cash}), (\ref{it:risk-convex}), and (\ref{it:risk-fatou}) below. A risk measure $\varrho$ is said to have the Lebesgue property if it satisfies (\ref{it:risk-lebesgue}).
\begin{enumerate}[label=\upshape\Roman*.,ref=\upshape\Roman*]
    \item\label{it:risk-monotone} Monotonicity: for each $\mathcal{X},\mathcal{Y}\in L^{\infty}(\mathbb{P})$, if $\mathcal{X}\geq\mathcal{Y}$, then $\varrho(\mathcal{X})\geq\varrho(\mathcal{Y})$.
    \item\label{it:risk-cash} Cash additivity: for each $\mathcal{X}\in L^{\infty}(\mathbb{P})$, if $c\in\mathbb{R}$, then $\varrho(\mathcal{X}+c)=\varrho(\mathcal{X})+c$.
    \item\label{it:risk-convex} Convexity: for each $\mathcal{X},\mathcal{Y}\in L^{\infty}(\mathbb{P})$ and $\lambda\in[0,1]$, $\varrho(\lambda\mathcal{X}+(1-\lambda)\mathcal{Y})\leq\lambda\varrho(\mathcal{X})+(1-\lambda)\varrho(\mathcal{Y})$.
    \item\label{it:risk-fatou} Fatou property: if $(\mathcal{X}^{n})_{n=1}^{\infty}\subseteq L^{\infty}(\mathbb{P})$ is an $L^{\infty}(\mathbb{P})$-bounded sequence converging in probability to $\mathcal{X}\in L^{\infty}(\mathbb{P})$, then
    \begin{equation*}
        \varrho(\mathcal{X})\leq\liminf_{n\to\infty}\varrho(\mathcal{X}^{n}).
    \end{equation*}
    \item\label{it:risk-lebesgue} Lebesgue property: if $(\mathcal{X}^{n})_{n=1}^{\infty}\subseteq L^{\infty}(\mathbb{P})$ is an $L^{\infty}(\mathbb{P})$-bounded sequence converging in probability to $\mathcal{X}\in L^{\infty}(\mathbb{P})$, then $\lim_{n\to\infty}\varrho(\mathcal{X}^{n})$ exists and equals $\varrho(\mathcal{X})$.
\end{enumerate}
\end{definition}
Risk measures are often defined to only satisfy (\ref{it:risk-monotone}) and (\ref{it:risk-cash}) from Definition \ref{def:risk-measure}. The definition we use here is stronger, requiring convexity (\ref{it:risk-convex}) and the Fatou property (\ref{it:risk-fatou}) in addition to (\ref{it:risk-monotone}) and (\ref{it:risk-cash}). In principle, our basic framework does not rely on any of the above axioms, although essentially all of our substantive results depend on some combination of (\ref{it:risk-monotone}) through (\ref{it:risk-fatou}) being satisfied.
\par
If $\varrho$ is a risk measure, we have the dual representation
\begin{equation}
    \varrho(\mathcal{X})=\sup_{\mathbb{Q}\in\mathscr{M}_{\mathbb{P}}}\left(\mathbb{E}^{\mathbb{Q}}(\mathcal{X})-\varrho^{\ast}(\mathbb{Q})\right),
\end{equation}
for each $\mathcal{X}\in L^{\infty}(\mathbb{P})$, where $\varrho^{\ast}(\mathbb{Q})=\sup_{\mathcal{X}\in L^{\infty}(\mathbb{P})}\left(\mathbb{E}^{\mathbb{Q}}(\mathcal{X})-\varrho(X)\right)$ for each $\mathbb{Q}\in\mathscr{M}_{\mathbb{P}}$ (see \cite[Theorem 2]{melnikov-duality}). The function $\varrho^{\ast}$ is called the convex conjugate of $\varrho$, and is well-defined even if $\varrho$ is not a risk measure. If $\varrho^{\ast}(\mathscr{M}_{\mathbb{P}})\subseteq\{0,\infty\}$, we say that $\varrho$ is coherent.
\par
For each agent $a\in A$, we therefore have a risk measure $\varrho_{a}$, codifying the risk preferences of agent $a$. Collecting all of the preferences yields a collection $(\varrho_{a})_{a\in A}$ of risk measures.
\par
Consider now an $\mathcal{X}$-feasible allocation $(X_{a})_{a\in A}$. The goal of risk sharing is to minimize some measure of total risk $\mathrm{TR}$. Translating the formulas from the discrete case into the language of integration yields a formula of the form
\begin{equation*}
    \mathrm{TR}=\int_{A}\varrho_{a}(X_{a})\mu(da).
\end{equation*}
Unfortunately, the above integral need not be well-defined—it is unclear that the real-valued function $a\longmapsto\varrho_{a}(X_{a})$ is measurable or integrable. The integrability issue is settled by setting $\int_{A}\varrho_{a}(X_{a})\mu(da)=\infty$ whenever $a\longmapsto\varrho_{a}(X_{a})$ is measurable and $\int_{A}\max\left\{0,\varrho_{a}(X_{a})\right\}\mu(da)=\infty$. The measurability issue is resolved by restricting the possible collections of preferences $(\varrho_{a})_{a\in A}$ to those that satisfy the following definition.
\begin{definition}\label{def:a-meas-pref}
    An indexed collection $(\varrho_{a})_{a\in A}$ of risk measures is said to be $\mathscr{A}$-measurable if, for each $\mathscr{A}$-measurable allocation $(X_{a})_{a\in A}$, the real-valued function $a\longmapsto\varrho_{a}(X_{a})$ is $\mathscr{A}$-measurable.
\end{definition}
\begin{example}\label{ex:single-fixed}
    Since $(\Omega,\mathscr{F},\mathbb{P})$ is a separable probability space, the collection $(\varrho_{a})_{a\in A}$ defined by setting $\varrho_{a}=\varrho$ for all $a\in A$ for some fixed risk measure $\varrho$ is $\mathscr{A}$-measurable.\footnote{Indeed, every convex lower $\sigma(L^{\infty},L^{1})$-semicontinuous function $\varphi$ can be represented as a supremum
    \begin{equation*}
        \varphi(\mathcal{X})=\sup_{(\mathcal{Y},b)\in C}\left(\mathbb{E}^{\mathbb{P}}\left(\mathcal{Y}\mathcal{X}\right)+b\right)
    \end{equation*}
    for all $\mathcal{X}\in L^{\infty}(\mathbb{P})$, where $C\subseteq L^{1}(\mathbb{P})\oplus\mathbb{R}$ is a set of $\sigma(L^{\infty},L^{1})$-continuous affine functions, which one can replace by a countable dense subset $C'\subseteq C$ by virtue of separability, yielding measurability.}
    \qed
\end{example}
\begin{example}\label{ex:multiple-risks}
    As a consequence of the conclusion of Example \ref{ex:single-fixed}, the collection $(\varrho_{a})_{a\in A}$ defined by
    \begin{equation*}
        a\longmapsto\varrho_{a}=\sum_{i=1}^{N}\mathbf{1}_{B_{i}}(a)\varrho_{i}
    \end{equation*}
    is $\mathscr{A}$-measurable for risk measures $(\varrho_{i})_{i=1}^{N}$ and a disjoint $\mathscr{A}$-measurable partition $(B_{i})_{i=1}^{N}$ of $A$.\qed
\end{example}
Since many other preferences are simply a limiting case of Example \ref{ex:multiple-risks}, $\mathscr{A}$-measurability of $(\varrho_{a})_{a\in A}$ is not a stringent condition (see, in particular, Theorem \ref{thm:dilate-share} and Theorem \ref{thm:inflate-dual} below).
\subsection{The Risk Sharing Problem}
Consider an element $\mathcal{X}\in L^{\infty}(\mathbb{P})$, to be allocated by a social planner among the agents in $A$, and fix an $\mathscr{A}$-measurable collection $(\varrho_{a})_{a\in A}$ of risk measures. The goal of the social planner is
\begin{equation*}
    \textrm{minimize }\int_{A}\varrho_{a}(X_{a})\mu(da)\textrm{, subject to $(X_{a})_{a\in A}\in C$}
\end{equation*}
where $C\subseteq \mathbb{A}(\mathcal{X})$ is a subset of the $\mathcal{X}$-feasible allocations. The subset $C$ can be strict, corresponding to a constrained version of the risk sharing problem, or $C$ may equal all of $\mathbb{A}(\mathcal{X})$, corresponding to an unconstrained version of the risk-sharing problem. Some commonly imposed constraints include individual rationality and comonotonicity of allocations (see \cite{ghossoubzhu2025,comon-markets}). Notwithstanding the importance of these directions, in this article we consider only the unconstrained case.
\section{The Unconstrained Value Function}\label{sec:unconstrain}
In this section, we study the value function of the risk sharing problem, which we introduce in \S\ref{subsec:val-function}. Our main result, Theorem \ref{thm:dual-rep-represent}, is contained in \S\ref{subsec:dual-val}. Theorem \ref{thm:dual-rep-represent} characterizes the convex conjugate of the value function, expressing it in terms of the convex conjugates $(\varrho^{\ast}_{a})_{a\in A}$.
\subsection{An Integral Infimal Convolution}\label{subsec:val-function}
Given risk measures $\varrho_{1}$ and $\varrho_{2}$, their infimal convolution $\varrho_{1}\operatorname{\Box}\varrho_{2}$ is defined by
\begin{equation*}
    \left(\varrho_{1}\operatorname{\Box}\varrho_{2}\right)(\mathcal{X})=\inf_{\mathcal{Y}\in L^{\infty}(\mathbb{P})}\left(\varrho_{1}(\mathcal{Y})+\varrho_{2}(\mathcal{X}-\mathcal{Y})\right)
\end{equation*}
for any $\mathcal{X}\in L^{\infty}(\mathbb{P})$ (see \cite[\S4.1]{quantile-share}). Naturally, the above definition can be extended to form the infimal convolution $\operatorname{\Box}_{i=1}^{N}\varrho_{i}$ of a finite sequence $(\varrho_{i})_{i=1}^{N}$ of risk measures. Motivated by this definition, we define the integral infimal convolution, generalizing the classical concept, as follows.
\begin{definition}
    If $(\varrho_{a})_{a\in A}$ is an $\mathscr{A}$-measurable collection of risk measures, the integral infimal convolution $\operatorname{\Box}_{a\in A}\varrho_{a}\mu(da)$ of $(\varrho_{a})_{a\in A}$ is defined as
    \begin{equation*}
        \left(\operatorname{\Box}_{a\in A}\varrho_{a}\mu(da)\right)(\mathcal{X})=\inf_{(X_{a})_{a\in A}\in\mathbb{A}(\mathcal{X})}\int_{A}\varrho_{a}(X_{a})\mu(da)
    \end{equation*}
    for each $\mathcal{X}\in L^{\infty}(\mathbb{P})$.
\end{definition}
Compared to the infimal convolution $\operatorname{\Box}_{i=1}^{N}\varrho_{i}$ of a finite sequence $(\varrho_{i})_{i=1}^{N}$ of risk measures, the integral infimal convolution may present some differences even when $A=\{1,\dots,N\}$, as in Example \ref{ex:finite}. Indeed, the measure $\mu$ functions as a weighting scheme implicit in all calculations, and so cannot in general be discarded. Viewed this way, the integral infimal convolution generalizes some of the existing weighting schemes for the infimal convolution (see \cite{righi-op}).
\par
It is not clear that the integral infimal convolution takes finite values. A sufficient condition for finiteness is partial agreement on priors, as we now show. Such assumptions have appeared in the literature before to guarantee finiteness of the value function (see \cite[Condition (E)]{lpconve}).
\begin{proposition}\label{prop:glob-finite}
    Suppose the $\mathscr{A}$-measurable collection $(\varrho_{a})_{a\in A}$ of risk measures consists of risk measures with the Lebesgue property, there exists $\mathbb{Q}\in\bigcap_{a\in A}\left\{\varrho^{\ast}_{a}<\infty\right\}$ such that $\int_{A}\varrho^{\ast}_{a}(\mathbb{Q})\mu(da)<\infty$, and $\int_{A}\vert{\varrho_{a}(0)\vert}\mu(da)<\infty$. Then $\operatorname{\Box}_{a\in A}\varrho_{a}\mu(da)$ is globally finite: for each $\mathcal{X}\in L^{\infty}(\mathbb{P})$, we have that $-\infty<\left(\operatorname{\Box}_{a\in A}\varrho_{a}\mu(da)\right)(\mathcal{X})<\infty$.
\end{proposition}
Note that, by Lemma \ref{lem:sup-measurable} in Appendix \ref{sec:tech}, Lemma \ref{lem:dual-accept-set} in Appendix \ref{sec:prf-of-thm-rep}, and Lemma \ref{lem:measurable-corr-accept} in Appendix \ref{sec:prf-of-thm-rep} (which assumes the Lebesgue property), the function $a\longmapsto\varrho^{\ast}_{a}(\mathbb{Q})$ is $\mathscr{A}$-measurable for each $\mathbb{Q}\in\mathscr{M}_{\mathbb{P}}$. Thus, the integral $\int_{A}\varrho^{\ast}_{a}(\mathbb{Q})\mu(da)$ is well-defined for each $\mathbb{Q}\in\mathscr{M}_{\mathbb{P}}$ in the context of Proposition \ref{prop:glob-finite}. At no other point in the proof of Proposition \ref{prop:glob-finite} do we use the Lebesgue property.
\begin{proof}
    We first show that $\left(\operatorname{\Box}_{a\in A}\varrho_{a}\mu(da)\right)(\mathcal{X})<\infty$ for all $\mathcal{X}\in L^{\infty}(\mathbb{P})$. Fixing an arbitrary $\mathcal{X}\in L^{\infty}(\mathbb{P})$, define $(X_{a})_{a\in A}\in\mathbb{A}(\mathcal{X})$ by $X_{a}=\mathcal{X}/\mu(A)$. Cash additivity implies
    \begin{equation*}
        \left(\operatorname{\Box}_{a\in A}\varrho_{a}\mu(da)\right)(\mathcal{X})\leq\int_{A}\varrho_{a}(X_{a})\mu(da)\leq\int_{A}\left(\vert\varrho_{a}(0)\vert+\Vert{\mathcal{X}}\Vert_{L^{\infty}}/\mu(A)\right)\mu(da)<\infty,
    \end{equation*}
    showing that $\left(\operatorname{\Box}_{a\in A}\varrho_{a}\mu(da)\right)(\mathcal{X})<\infty$. Since $\mathcal{X}\in L^{\infty}(\mathbb{P})$ was arbitrary, this proves the claim.
    \par
    We now show that $\left(\operatorname{\Box}_{a\in A}\varrho_{a}\mu(da)\right)(\mathcal{X})>-\infty$ for all $\mathcal{X}\in L^{\infty}(\mathbb{P})$. Fix an arbitrary $\mathcal{X}\in L^{\infty}(\mathbb{P})$, and $\mathbb{Q}\in\bigcap_{a\in A}\left\{\varrho^{\ast}_{a}<\infty\right\}$ such that $\int_{A}\varrho^{\ast}_{a}(\mathbb{Q})\mu(da)<\infty$. Then, for any $(X_{a})_{a\in A}\in\mathbb{A}(\mathcal{X})$,
    \begin{equation*}
        \int_{A}\varrho_{a}(X_{a})\mu(da)\geq\int_{A}\left(\mathbb{E}^{\mathbb{Q}}\left(X_{a}\right)-\varrho^{\ast}_{a}(\mathbb{Q})\right)\mu(da)=\mathbb{E}^{\mathbb{Q}}\left(\mathcal{X}\right)-\int_{A}\varrho^{\ast}_{a}(\mathbb{Q})\mu(da).
    \end{equation*}
    Thus, taking the infimum over $(X_{a})_{a\in A}\in\mathbb{A}(\mathcal{X})$,
    \begin{equation*}
        \left(\operatorname{\Box}_{a\in A}\varrho_{a}\mu(da)\right)(\mathcal{X})\geq\mathbb{E}^{\mathbb{Q}}\left(\mathcal{X}\right)-\int_{A}\varrho^{\ast}_{a}(\mathbb{Q})\mu(da)>-\infty.
    \end{equation*}
    Since $\mathcal{X}\in L^{\infty}(\mathbb{P})$ was arbitrary, this proves the claim.
\end{proof}
Even when the integral infimal convolution is globally finite, it is not necessarily a risk measure, as it may fail to satisfy the Fatou property (this occurs even in the two agent case—see \cite[Theorem 12]{mackey-constraints}). Convex duality, on which our later results hinge, can be applied to functionals failing the Fatou property, but the relevant dual pairing will include non-$\sigma$-additive set functions, which are here squarely foreclosed from consideration by the weakness of the measurability assumptions we make. Thus, compelled by necessity, we now consider sufficient conditions for the integral infimal convolution to possess the Fatou property.
\begin{proposition}\label{prop:value-funct-is-fatou}
    Suppose the $\mathscr{A}$-measurable collection $(\varrho_{a})_{a\in A}$ of risk measures consists of risk measures with the Lebesgue property. Then, if $\operatorname{\Box}_{a\in A}\varrho_{a}\mu(da)$ is globally finite, $\operatorname{\Box}_{a\in A}\varrho_{a}\mu(da)$ is a risk measure with the Lebesgue property (hence, a fortiori, the Fatou property holds).
\end{proposition}
\begin{proof}
    Monotonicity, convexity, and cash additivity are not difficult to prove, and we therefore focus on the Lebesgue property. It suffices to prove continuity from above (see Remark \ref{rem:cont-from-above}), in the sense that if $(\mathcal{X}^{n})_{n=1}^{\infty}\subseteq L^{\infty}(\mathbb{P})$ is decreasing and converges $\mathbb{P}$-a.s. to $\mathcal{X}\in L^{\infty}(\mathbb{P})$, then
    \begin{equation*}
        \inf_{n}\left(\operatorname{\Box}_{a\in A}\varrho_{a}\mu(da)\right)(\mathcal{X}^{n})=\left(\operatorname{\Box}_{a\in A}\varrho_{a}\mu(da)\right)(\mathcal{X}).
    \end{equation*}
    To this end, note that
    \begin{equation*}
        \inf_{n}\left(\operatorname{\Box}_{a\in A}\varrho_{a}\mu(da)\right)(\mathcal{X}^{n})=\inf_{n}\inf_{(X_{a})_{a\in A}\in\mathbb{A}(0)}\int_{A}\varrho_{a}(\mathcal{X}^{n}/\mu(A)+X_{a})\mu(da)
    \end{equation*}
    \begin{equation*}
        =\inf_{(X_{a})_{a\in A}\in\mathbb{A}(0)}\inf_{n}\int_{A}\varrho_{a}(\mathcal{X}^{n}/\mu(A)+X_{a})\mu(da)=\inf_{(X_{a})_{a\in A}\in\mathbb{A}(0)}\int_{A}\inf_{n}\varrho_{a}(\mathcal{X}^{n}/\mu(A)+X_{a})\mu(da)
    \end{equation*}
    \begin{equation*}
        =\inf_{(X_{a})_{a\in A}\in\mathbb{A}(0)}\int_{A}\varrho_{a}(\mathcal{X}/\mu(A)+X_{a})\mu(da)=\left(\operatorname{\Box}_{a\in A}\varrho_{a}\mu(da)\right)(\mathcal{X})
    \end{equation*}
    by the monotone convergence theorem and the Lebesgue property of each $\varrho_{a}$, establishing the claim.
\end{proof}
\begin{remark}\label{rem:cont-from-above}
    Remark 4.19 of Föllmer and Schied \cite{stoch-fin} allows us to deduce the Lebesgue property from continuity from above in the proof of Proposition \ref{prop:value-funct-is-fatou}. Since they use a different sign convention, we provide a brief explanation. Denoting by $M_{-1}$ the map $\mathcal{X}\longmapsto-\mathcal{X}$, a risk measure $\varrho$ in our framework can be converted into a risk measure in the sense of Föllmer and Schied by defining $\widehat{\varrho}=\varrho\circ M_{-1}$. The Lebesgue property of $\varrho$ is equivalent to that of $\widehat{\varrho}$. By contrast, continuity from above of $\varrho$ is equivalent to continuity from below of $\widehat{\varrho}$. In view of \cite[Remark 4.19]{stoch-fin}, $\widehat{\varrho}$ has the Lebesgue property whenever $\widehat{\varrho}$ is continuous from below. Hence $\varrho$ has the Lebesgue property whenever $\varrho$ is continuous from above.
    \qed
\end{remark}
\subsection{Dual Representations}\label{subsec:dual-val}
Given a finite sequence $(\varrho_{i})_{i=1}^{N}$ of risk measures, it is known (see \cite[Remark 3.105]{BarbuPrecupanu2012}) that their infimal convolution $\operatorname{\Box}_{i=1}^{N}\varrho_{i}$ satisfies
\begin{equation}\label{eq:classical-dual-rep-inf}
    \left(\operatorname{\Box}_{i=1}^{N}\varrho_{i}\right)^{\ast}=\sum_{i=1}^{N}\varrho^{\ast}_{i}.
\end{equation}
One can generalize this fact to the integral infimal convolution defined in the previous subsection. The formula remains essentially the same, with the finite sum in (\ref{eq:classical-dual-rep-inf}) replaced by an integral.
\begin{theorem}\label{thm:dual-rep-represent}
    Suppose the $\mathscr{A}$-measurable collection $(\varrho_{a})_{a\in A}$ of risk measures consists of risk measures with the Lebesgue property, $\int_{A}\vert{\varrho_{a}(0)}\vert\mu(da)<\infty$, and $\operatorname{\Box}_{a\in A}\varrho_{a}\mu(da)$ is globally finite. Then, for each $\mathbb{Q}\in\mathscr{M}_{\mathbb{P}}$, the function $a\longmapsto\varrho^{\ast}_{a}(\mathbb{Q})$ is $\mathscr{A}$-measurable, and
    \begin{equation}\label{eq:conj-of-inf}
        \left(\operatorname{\Box}_{a\in A}\varrho_{a}\mu(da)\right)^{\ast}(\mathbb{Q})=\int_{A}\varrho^{\ast}_{a}(\mathbb{Q})\mu(da).
    \end{equation}
    Furthermore,
    \begin{equation}\label{eq:dual-of-inf}
        \left(\operatorname{\Box}_{a\in A}\varrho_{a}\mu(da)\right)(\mathcal{X})=\sup_{\mathbb{Q}\in\mathscr{M}_{\mathbb{P}}}\left(\mathbb{E}^{\mathbb{Q}}(\mathcal{X})-\int_{A}\varrho^{\ast}_{a}(\mathbb{Q})\mu(da)\right).
    \end{equation}
\end{theorem}
\begin{proof}
    See Appendix \ref{sec:prf-of-thm-rep} for the proof of (\ref{eq:conj-of-inf}). By Proposition \ref{prop:value-funct-is-fatou}, the integral infimal convolution is a risk measure under the assumptions of Theorem \ref{thm:dual-rep-represent}, and therefore (\ref{eq:dual-of-inf}) holds in light of (\ref{eq:conj-of-inf}).
\end{proof}
Remark that, while the conjugate of the value function looks the same as the finite agent case, behavior related to optimizers of the risk sharing problem can differ under a continuum of agents (see \S\ref{subsec:inflate}).
\par
In the process of proving Theorem \ref{thm:dual-rep-represent}, some interesting complementary results are derived in Appendix \ref{sec:prf-of-thm-rep}. In particular, we characterize random variables with non-positive integral infimal convolution via Aumann integration (see Theorem \ref{thm:accept-set-characterize}), generalizing earlier results which used Minkowski summation (see the proof of \cite[Theorem 4.1]{felixfs}).
\subsection{A First Order Optimality Condition}
Theorem \ref{thm:dual-rep-represent} has consequences for the behavior of optimizers: they must share a common subgradient. Recall that a subgradient (see \cite[Definition 2.30]{BarbuPrecupanu2012}) of a risk measure $\varrho$ is a point $\mathbb{Q}\in\mathscr{M}_{\mathbb{P}}$ such that
\begin{equation*}
    \forall\mathcal{Y}\in L^{\infty}(\mathbb{P}):\varrho(\mathcal{Y})-\varrho(\mathcal{X})\geq\mathbb{E}^{\mathbb{Q}}(\mathcal{Y}-\mathcal{X}),
\end{equation*}
with the set of subgradients of $\varrho$ at $\mathcal{X}$ denoted by $\partial\varrho(\mathcal{X})$. Equivalently, $\mathbb{Q}\in\partial\varrho(\mathcal{X})$ iff $\varrho(\mathcal{X})=\mathbb{E}^{\mathbb{Q}}(\mathcal{X})-\varrho^{\ast}(\mathbb{Q})$ (see \cite[Proposition 2.33]{BarbuPrecupanu2012}).
\begin{proposition}\label{prop:subdifferential}
    Let $\mathbb{Q}\in\mathscr{M}_{\mathbb{P}}$ be any subgradient of the integral infimal convolution of $(\varrho_{a})_{a\in A}$ at $\mathcal{X}$, and let $(X_{a})_{a\in A}\in\mathbb{A}(\mathcal{X})$. If the assumptions of Theorem \ref{thm:dual-rep-represent} are satisfied, then
    \begin{equation*}
        \mathbb{Q}\in\partial\varrho_{a}(X_{a})
    \end{equation*}
    for $\mu$-a.e. $a\in A$ if, and only if, $(X_{a})_{a\in A}$ is an optimizer of the integral infimal convolution of $(\varrho_{a})_{a\in A}$.
\end{proposition}
In the context of Proposition \ref{prop:subdifferential}, a subgradient of the integral infimal convolution of $(\varrho_{a})_{a\in A}$ at $\mathcal{X}$ always exists, a consequence of Proposition \ref{prop:value-funct-is-fatou} conjoined with the Jouini-Schachermayer-Touzi theorem (see \cite[Theorem 2.4]{owari-jfa}).
\begin{proof}
    Suppose $(X_{a})_{a\in A}$ is optimal. If $\mathbb{Q}$ is a subgradient of $\operatorname{\Box}_{a\in A}\varrho_{a}\mu(da)$ at $\mathcal{X}$, we have that
    \begin{equation*}
        \left(\operatorname{\Box}_{a\in A}\varrho_{a}\mu(da)\right)(\mathcal{X})=\mathbb{E}^{\mathbb{Q}}(\mathcal{X})+\left(\operatorname{\Box}_{a\in A}\varrho_{a}\mu(da)\right)^{\ast}(\mathbb{Q}).
    \end{equation*}
    After applying Theorem \ref{thm:dual-rep-represent} and using the fact that $(X_{a})_{a\in A}$ is an optimizer, we get that
    \begin{equation}\label{eq:sub-in-optimal}
        \int_{A}\varrho_{a}(X_{a})\mu(da)=\mathbb{E}^{\mathbb{Q}}(\mathcal{X})-\int_{A}\varrho^{\ast}_{a}(\mathbb{Q})\mu(da)=\int_{A}\left(\mathbb{E}^{\mathbb{Q}}(X_{a})-\varrho^{\ast}_{a}(\mathbb{Q})\right)\mu(da).
    \end{equation}
    By Fenchel's inequality, $\varrho_{a}(X_{a})\geq\mathbb{E}^{\mathbb{Q}}(X_{a})-\varrho^{\ast}_{a}(\mathbb{Q})$ for all $a\in A$; if this inequality is strict on a set of non-zero $\mu$-measure, (\ref{eq:sub-in-optimal}) must fail. So $\varrho_{a}(X_{a})=\mathbb{E}^{\mathbb{Q}}(X_{a})-\varrho^{\ast}_{a}(\mathbb{Q})$ for $\mu$-a.e. $a\in A$. In particular, $\mathbb{Q}\in\partial\varrho_{a}(X_{a})$ for $\mu$-a.e. $a\in A$, as desired.
    \par
    For the converse, suppose $\mathbb{Q}\in\partial\varrho_{a}(X_{a})$ for $\mu$-a.e. $a\in A$. Then
    \begin{equation*}
        \left(\operatorname{\Box}_{a\in A}\varrho_{a}\mu(da)\right)(\mathcal{X})=\mathbb{E}^{\mathbb{Q}}(\mathcal{X})-\int_{A}\varrho^{\ast}_{a}(\mathbb{Q})\mu(da)
    \end{equation*}
    \begin{equation*}
        =\int_{A}\left(\mathbb{E}^{\mathbb{Q}}(X_{a})-\varrho^{\ast}_{a}(\mathbb{Q})\right)\mu(da)=\int_{A}\varrho_{a}(X_{a})\mu(da)
    \end{equation*}
    in light of Theorem \ref{thm:dual-rep-represent}, so $(X_{a})_{a\in A}$ is optimal.
\end{proof}
\section{Examples}\label{sec:examples}
We now apply the theory developed in \S\ref{sec:unconstrain} to specific families of risk measures, including those that fall in the class of entropic or expected shortfall risk measures.
\par
In \S\ref{subsec:dilate}, dilations of a fixed risk measure $\varrho$ are considered, and explicit formulas for the value function and optimal allocation are given (see Theorem \ref{thm:dilate-share}). Risk preferences modeled by entropic risk measures at various risk tolerance levels are covered by the results of this subsection.
\par
In \S\ref{subsec:inflate}, we define a family of risk measures obtained from a fixed coherent risk measure $\varrho$, which we call inflations of $\varrho$. An explicit formula is given for the value function when risk preferences are inflations of a fixed coherent risk measure (see Theorem \ref{thm:inflate-dual}), and sufficient conditions are given for the existence and non-existence of an optimal allocation (see Theorem \ref{thm:attain-notattain} and Appendix \ref{sec:prf-attain-notattain}). Risk preferences modeled by expected shortfall at various quantile levels are covered by the results of this subsection.
\subsection{Dilated Risk Measures}\label{subsec:dilate}
For a risk measure $\varrho$, Barrieu and El Karoui \cite{opt-risk-transfer} introduced the family $(\varrho^{\gamma})_{\gamma\in(0,\infty)}$ of dilated risk measures (see \cite[Definition 3.8]{opt-risk-transfer}).
\begin{definition}
    Let $\varrho$ be a risk measure, and fix $\gamma>0$. The $\gamma$-dilation $\varrho^{\gamma}$ of $\varrho$ is defined by
    \begin{equation*}
        \varrho^{\gamma}(\mathcal{X})=\gamma\varrho\left(\frac{1}{\gamma}\mathcal{X}\right)
    \end{equation*}
    for any $\mathcal{X}\in L^{\infty}(\mathbb{P})$.
\end{definition}
In some circumstances, dilation may fail to produce any new non-trivial risk measures, as the following example illustrates.
\begin{example}\label{ex:triv-dilation}
    Suppose $\varrho$ is a coherent risk measure, so that $\varrho^{\ast}(\mathscr{M}_{\mathbb{P}})\subseteq\{0,\infty\}$. Then, $\varrho^{\gamma}=\varrho$ for each $\gamma>0$.
    \qed
\end{example}
The triviality of Example \ref{ex:triv-dilation} is not universal; dilation can produce a non-trivial new family of risk measures, as demonstrated by the class of entropic risk measures.
\begin{example}
    For a risk tolerance parameter $\gamma>0$, the entropic risk measure $\mathrm{Ent}^{\gamma}$ is defined as
    \begin{equation*}
        \mathrm{Ent}^{\gamma}(\mathcal{X})=\gamma\log\left(\mathbb{E}^{\mathbb{P}}\left(e^{\frac{1}{\gamma} \mathcal{X}}\right)\right)=\sup_{\mathbb{Q}\in\mathscr{M}_{\mathbb{P}}}\left(\mathbb{E}^{\mathbb{Q}}(\mathcal{X})-\gamma H(\mathbb{Q}\parallel\mathbb{P})\right)
    \end{equation*}
    for any $\mathcal{X}\in L^{\infty}(\mathbb{P})$, where $H(\mathbb{Q}\parallel\mathbb{P})=\mathbb{E}^{\mathbb{Q}}\left(\log\left(\frac{d\mathbb{Q}}{d\mathbb{P}}\right)\right)$ is the relative entropy of $\mathbb{Q}\ll\mathbb{P}$. It is easy to see that $\mathrm{Ent}^{\gamma}$ is the $\gamma$-dilation of $\mathrm{Ent}^{1}$, and that $\mathrm{Ent}^{\gamma}\neq\mathrm{Ent}^{\gamma'}$ for $\gamma\neq\gamma'$ whenever $(\Omega,\mathscr{F},\mathbb{P})$ is sufficiently non-trivial.
    \qed
\end{example}
We now state the main result of this section, which explicitly derives the value function and optimal allocation when risk preferences are dilations of a fixed risk measure. Of particular note is the fact that an optimal allocation always exists, and an explicit formula is given. This explicit formula generalizes results from the discrete case (for instance, those of Righi and Moresco \cite{righi-op}).
\begin{theorem}\label{thm:dilate-share}
    Let $\varrho$ be a risk measure with the Lebesgue property, and let $(\gamma_{a})_{a\in A}\in(0,\infty)^{A}$ be an $\mathscr{A}$-measurable map with $\int_{A}\gamma_{a}\mu(da)<\infty$. Defining $\varrho_{a}=\varrho^{\gamma_{a}}$ for each $a\in A$ and $\Gamma=\int_{A}\gamma_{a}\mu(da)$, we have the following.
    \begin{enumerate}
        \item\label{item:inflate-measurable} The indexed collection $(\varrho_{a})_{a\in A}$ of risk measures is $\mathscr{A}$-measurable.
        \item\label{item:inflate-convolution} The integral infimal convolution $\operatorname{\Box}_{a\in A}\varrho_{a}\mu(da)$ satisfies
        \begin{equation*}
            \operatorname{\Box}_{a\in A}\varrho_{a}\mu(da)=\varrho^{\Gamma}.
        \end{equation*}
        In particular,
        \begin{equation*}
            \left(\operatorname{\Box}_{a\in A}\varrho_{a}\mu(da)\right)^{\ast}=\Gamma\varrho^{\ast}.
        \end{equation*}
        \item\label{item:inflate-optimal} For any $\mathcal{X}\in L^{\infty}(\mathbb{P})$, the allocation $(\gamma_{a}\mathcal{X}/\Gamma)_{a\in A}\in\mathbb{A}(\mathcal{X})$ is optimal, in the sense that
        \begin{equation*}
            \left(\operatorname{\Box}_{a\in A}\varrho_{a}\mu(da)\right)(\mathcal{X})=\int_{A}\varrho_{a}(\gamma_{a}\mathcal{X}/\Gamma)\mu(da).
        \end{equation*}
    \end{enumerate}
\end{theorem}
\begin{proof}
    For (1), we may find a sequence $\left((\gamma^{n}_{a})_{a\in A}\right)_{n=1}^{\infty}$ of $\mathscr{A}$-measurable simple functions, such that $\lim_{n\to\infty}\gamma^{n}=\gamma$ pointwise. By replacing $\gamma^{n}$ with $\gamma^{n}\vee\frac{1}{n}$ if necessary, we may assume that $\gamma^{n}$ takes values in $(0,\infty)$. By the argument in Example \ref{ex:multiple-risks}, for each $n$, the family $(\varrho^{n}_{a})_{a\in A}$ of risk measures defined by $\varrho^{n}_{a}=\varrho^{\gamma^{n}_{a}}$ for each $a\in A$ is $\mathscr{A}$-measurable. Thus, for each $\mathscr{A}$-measurable allocation $(X_{a})_{a\in A}$, $a\longmapsto\varrho^{n}_{a}(X_{a})$ is $\mathscr{A}$-measurable. As $n\to\infty$, the Lebesgue property of $\varrho$ implies that $\lim_{n\to\infty}\varrho^{n}_{a}(X_{a})$ exists and equals $\varrho_{a}(X_{a})$. Since pointwise limits of $\mathscr{A}$-measurable functions are $\mathscr{A}$-measurable, this implies that $a\longmapsto\varrho_{a}(X_{a})$ is $\mathscr{A}$-measurable. Since $(X_{a})_{a\in A}$ was an arbitrary $\mathscr{A}$-measurable allocation, this shows that $(\varrho_{a})_{a\in A}$ is an $\mathscr{A}$-measurable collection of risk measures, proving (1).
    \par
    To establish (2), we use Theorem \ref{thm:dual-rep-represent}. First, one must show that the preconditions for Theorem \ref{thm:dual-rep-represent} hold. Thus, one must establish the following:
    \begin{enumerate}[label=\upshape\roman*.,ref=\upshape\roman*]
        \item\label{item:roman-inflate-integral-finite} $\int_{A}\vert{\varrho_{a}}(0)\vert\mu(da)<\infty$.
        \item\label{item:roman-inflate-lebesgue} The collection $(\varrho_{a})_{a\in A}$ consists of risk measures with the Lebesgue property.
        \item\label{item:roman-inflate-globfinite} The integral infimal convolution $\operatorname{\Box}_{a\in A}\varrho_{a}\mu(da)$ is globally finite.
    \end{enumerate}
    By the definition of dilation,
    \begin{equation*}
        \int_{A}\vert{\varrho_{a}(0)\vert}\mu(da)=\int_{A}\vert{\gamma_{a}\varrho(0)\vert}\mu(da)=\Gamma\vert{\varrho(0)}\vert<\infty,
    \end{equation*}
    implying (\ref*{item:roman-inflate-integral-finite}). Since $\varrho$ has the Lebesgue property, and all dilations of $\varrho$ therefore have the Lebesgue property, (\ref*{item:roman-inflate-lebesgue}) holds. To establish (\ref*{item:roman-inflate-globfinite}), it suffices to verify the preconditions of Proposition \ref{prop:glob-finite}; (\ref*{item:roman-inflate-integral-finite}) and (\ref*{item:roman-inflate-lebesgue}) are both preconditions (both of which we have already verified), and the only remaining precondition is the existence of $\mathbb{Q}\in\bigcap_{a\in A}\{\varrho^{\ast}_{a}<\infty\}$ with $\int_{A}\varrho^{\ast}_{a}(\mathbb{Q})\mu(da)<\infty$. There exists $\mathbb{Q}\in\{\varrho^{\ast}<\infty\}$; since $\varrho^{\ast}_{a}(\mathbb{Q})=\gamma_{a}\varrho^{\ast}(\mathbb{Q})$, it follows that $\mathbb{Q}\in\bigcap_{a\in A}\{\varrho^{\ast}_{a}<\infty\}$. It is easy to see that
    \begin{equation*}
        \int_{A}\varrho^{\ast}_{a}(\mathbb{Q})\mu(da)=\int_{A}\gamma_{a}\varrho^{\ast}(\mathbb{Q})\mu(da)=\Gamma\varrho^{\ast}(\mathbb{Q})<\infty,
    \end{equation*}
    establishing (\ref*{item:roman-inflate-globfinite}).
    \par
    We now apply Theorem \ref{thm:dual-rep-represent}. By Theorem \ref{thm:dual-rep-represent}, for all $\mathbb{Q}\in\mathscr{M}_{\mathbb{P}}$,
    \begin{equation*}
        \left(\operatorname{\Box}_{a\in A}\varrho_{a}\mu(da)\right)^{\ast}(\mathbb{Q})=\int_{A}\varrho^{\ast}_{a}(\mathbb{Q})\mu(da)=\int_{A}\gamma_{a}\varrho^{\ast}(\mathbb{Q})\mu(da)=\Gamma\varrho^{\ast}(\mathbb{Q})
    \end{equation*}
    implying $\left(\operatorname{\Box}_{a\in A}\varrho_{a}\mu(da)\right)^{\ast}=\Gamma\varrho^{\ast}$, and hence also that $\operatorname{\Box}_{a\in A}\varrho_{a}\mu(da)=\varrho^{\Gamma}$. Thus, (2) holds.
    \par
    For (3), it suffices to show (using the explicit representation of the integral infimal convolution previously derived) that
    \begin{equation*}
        \varrho^{\Gamma}(\mathcal{X})=\int_{A}\varrho_{a}(\gamma_{a}\mathcal{X}/\Gamma)\mu(da).
    \end{equation*}
    By the definition of dilation,
    \begin{equation*}
        \int_{A}\varrho_{a}(\gamma_{a}\mathcal{X}/\Gamma)\mu(da)=\int_{A}\gamma_{a}\varrho(\mathcal{X}/\Gamma)\mu(da)=\Gamma\varrho(\mathcal{X}/\Gamma)=\varrho^{\Gamma}(\mathcal{X}),
    \end{equation*}
    as desired.
\end{proof}
As an illustration of the above result, we now apply Theorem \ref{thm:dilate-share} to the case of entropic risk measures.
\begin{example}\label{ex:entropic-share}
    Suppose $(\gamma_{a})_{a\in A}\in(0,\infty)^{A}$ is $\mathscr{A}$-measurable, and $\int_{A}\gamma_{a}\mu(da)<\infty$. Then, by virtue of Theorem \ref{thm:dilate-share}, the risk preferences $(\varrho_{a})_{a\in A}$ defined by $\varrho_{a}=\mathrm{Ent}^{\gamma_{a}}$ are such that
    \begin{equation*}
        \operatorname{\Box}_{a\in A}\varrho_{a}\mu(da)=\mathrm{Ent}^{\Gamma},
    \end{equation*}
    where $\Gamma=\int_{A}\gamma_{a}\mu(da)$. Thus, the integral infimal convolution of entropic risk measures is an entropic risk measure with the risk tolerance parameter defined by the total risk tolerance of agents in $A$.
    \par
    By Theorem \ref{thm:dilate-share}, for a given risk $\mathcal{X}\in L^{\infty}(\mathbb{P})$ to allocate, an optimal allocation of risk is $(\gamma_{a}\mathcal{X}/\Gamma)_{a\in A}$. Under this allocation, each agent $a\in A$ receives the portion of $\mathcal{X}$ defined by considering their proportion $\gamma_{a}/\Gamma$ of the total risk tolerance $\Gamma$. Also, by Proposition \ref{prop:subdifferential} and the characterization of subgradients of entropic risk measures in terms of Esscher tilting (see \cite[Lemma 6.1]{subgrad_law_inv}), the optimizer $(\gamma_{a}\mathcal{X}/\Gamma)_{a\in A}$ is unique, in the sense that any other optimizer $(X_{a})_{a\in A}\in\mathbb{A}(\mathcal{X})$ must take the form $X_{a}=C_{a}+\gamma_{a}\mathcal{X}/\Gamma$ $\mu$-a.e. for some cash-valued $(C_{a})_{a\in A}\in L^{1}(\mu)$ with $\int_{A}C_{a}\mu(da)=0$.
    \qed
\end{example}
\subsection{Inflated Risk Measures}\label{subsec:inflate}
In this subsection, we introduce a class of risk measures derived from a fixed coherent risk measure. Essentially, the class of probability measures for which the convex conjugate returns a finite value is enlarged. In \S\ref{subsubsec:inflate-properties}, we formally define inflated risk measures, explain their relation to worst-case expected shortfall measures, and obtain an analogue of the Rockafellar-Uryasev representation. The risk sharing problem for inflated risk measures is tackled in \S\ref{subsubsec:inflate-share}.
\subsubsection{Inflated Risk Measures: Basic Properties}\label{subsubsec:inflate-properties}
Let $\varrho$ be a coherent risk measure with $\varrho^{\ast}(\mathbb{P})=0$.\footnote{The latter assumption is not crucial to our analysis; we only invoke the assumption in one part of Proposition \ref{prop:inflate-continuous} and in Proposition \ref{prop:essup-and-inflation}. In fact, the role of $\mathbb{P}$ in those results can be replaced by any measure $\mathbb{Q}$ in the domain of $\varrho^{\ast}$ which majorizes every other element in the absolute continuity preorder; that such a majorant always exists follows from an exhaustion argument (see \cite[p. 77]{math-of-arb}).} Because $\varrho$ is coherent, $\varrho^{\ast}$ only takes values in the set $\{0,\infty\}$; thus, writing $\mathscr{Q}(\varrho)=\left\{\mathbb{Q}\in\mathscr{M}_{\mathbb{P}}:\varrho^{\ast}(\mathbb{Q})=0\right\}$, we have the dual representation
\begin{equation}\label{eq:dualsetrep}
    \varrho(\mathcal{X})=\sup_{\mathbb{Q}\in\mathscr{Q}(\varrho)}\mathbb{E}^{\mathbb{Q}}(\mathcal{X})
\end{equation}
for all $\mathcal{X}\in L^{\infty}(\mathbb{P})$. Define
\begin{equation*}
    \widetilde{\mathscr{Q}}(\varrho)=\left\{\mathcal{Y}\in L^{1}(\mathbb{P}):\exists\mathbb{Q}\in\mathscr{Q}(\varrho)\textrm{ such that }0\leq\mathcal{Y}\leq\frac{d\mathbb{Q}}{d\mathbb{P}}\right\}.
\end{equation*}
We abuse notation and identify, for any $\gamma\geq1$, $\gamma\widetilde{\mathscr{Q}}(\varrho)\cap\mathscr{M}_{\mathbb{P}}$ with the set of measures $\mathbb{Q}\in\mathscr{M}_{\mathbb{P}}$ such that $\frac{d\mathbb{Q}}{d\mathbb{P}}\in\gamma\widetilde{\mathscr{Q}}(\varrho)$.
\begin{definition}\label{def:inflate}
    Let $\varrho$ be a risk measure with dual representation (\ref{eq:dualsetrep}), and fix $\gamma\geq1$. The $\gamma$-inflation $\widetilde{\varrho}_{\gamma}$ of $\varrho$ is defined by
    \begin{equation*}
        \widetilde{\varrho}_{\gamma}(\mathcal{X})=\sup_{\mathbb{Q}\in\gamma\widetilde{\mathscr{Q}}(\varrho)\cap\mathscr{M}_{\mathbb{P}}}\mathbb{E}^{\mathbb{Q}}\left(\mathcal{X}\right)
    \end{equation*}
    for any $\mathcal{X}\in L^{\infty}(\mathbb{P})$.
\end{definition}
The idea of the above definition is that the set of probability measures used to calculate the dual representation is inflated by a factor of $\gamma$. There is another interpretation of inflated risk measures in terms of likelihood ratios, with the parameter $\gamma$ representing the model uncertainty faced by an agent. More precisely, $\mathbb{S}\in\gamma\widetilde{\mathscr{Q}}(\varrho)\cap\mathscr{M}_{\mathbb{P}}$ iff $\mathbb{S}\in\mathscr{M}_{\mathbb{P}}$ and there exists $\mathbb{Q}\in\mathscr{Q}(\varrho)$ with $\mathbb{S}\ll\mathbb{Q}$ and $\frac{d\mathbb{S}}{d\mathbb{Q}}\leq\gamma$. Hence if $\mathscr{Q}(\varrho)$ represents an agent's baseline set of priors, the $\gamma$-inflation enlarges the set of priors by including any $\mathbb{S}\in\mathscr{M}_{\mathbb{P}}$ whose likelihood ratio $\frac{d\mathbb{S}}{d\mathbb{Q}}$ is not too large for some prior $\mathbb{Q}\in\mathscr{Q}(\varrho)$. Likelihood ratios appear naturally here, as they are often used to quantify model uncertainty (see \cite{model-uncertainty-likelihood}).
\par
Some of the families generated by Definition \ref{def:inflate} are known, including expected shortfall.
\begin{example}
    For a quantile level $0<\alpha\leq1$, define the risk measure $\mathrm{ES}^{\alpha}$ by
    \begin{equation*}
        \mathrm{ES}^{\alpha}(\mathcal{X})=\sup_{\mathbb{Q}\in\mathscr{M}_{\mathbb{P}},\frac{d\mathbb{Q}}{d\mathbb{P}}\leq\frac{1}{\alpha}}\mathbb{E}^{\mathbb{Q}}(\mathcal{X})
    \end{equation*}
    for each $\mathcal{X}\in L^{\infty}(\mathbb{P})$. $\mathrm{ES}^{\alpha}$ is called the expected shortfall at quantile level $\alpha$. It is not difficult to see that $\mathrm{ES}^{\alpha}$ is the $\gamma$-inflation of $\mathrm{ES}^{1}=\mathbb{E}^{\mathbb{P}}$ for $\gamma=\frac{1}{\alpha}$.
    \qed
\end{example}
\begin{remark}
    If it is necessary to stress the underlying probability measure $\mathbb{P}$ from which expected shortfall is calculated, we will denote $\mathrm{ES}^{\alpha}=\mathrm{ES}^{\alpha}_{\mathbb{P}}$.
    \qed
\end{remark}
Expected shortfall measures are the canonical example of $\gamma$-inflation, as many properties of general inflated risk measures can be deduced from the corresponding properties of expected shortfall. The connection between expected shortfall and $\gamma$-inflation is also illustrated by the formula
\begin{equation*}
    \widetilde{\varrho}_{\gamma}=\sup_{\mathbb{Q}\in\mathscr{Q}(\varrho)}\mathrm{ES}^{q(\gamma)}_{\mathbb{Q}}
\end{equation*}
where $q(\gamma)=\frac{1}{\gamma}$ (see Proposition \ref{prop:inflate-es-relation}). In particular, inflated risk measures are worst-case expected shortfall measures in the sense of Zhu and Fukushima \cite{worst-case-es}, a widely studied class in the model uncertainty literature (see \cite[\S\S2.2,6.1]{partial-law-inv}). Conversely, a worst-case expected shortfall measure $\eta=\sup_{\mathbb{Q}\in\mathcal{Q}}\mathrm{ES}^{\alpha}_{\mathbb{Q}}$, calculated from a convex $L^{1}$-closed baseline set of priors $\mathcal{Q}\subseteq\mathscr{M}_{\mathbb{P}}$ containing $\mathbb{P}$, is a $\gamma$-inflation: $\eta=\widetilde{\varrho}_{\gamma}$ for $\varrho=\sup_{\mathbb{Q}\in\mathcal{Q}}\mathbb{E}^{\mathbb{Q}}$ and $\gamma=1/\alpha$.\footnote{The two notions are \textit{prima facie} distinct, because a general set $\mathcal{Q}\subseteq\mathscr{M}_{\mathbb{P}}$ need not be convex.}
\begin{proposition}\label{prop:inflate-es-relation}
    For each $\gamma\in[1,\infty)$,
    \begin{equation}\label{eq:es-rep-inflate}
        \widetilde{\varrho}_{\gamma}=\sup_{\mathbb{Q}\in\mathscr{Q}(\varrho)}\mathrm{ES}^{q(\gamma)}_{\mathbb{Q}}
    \end{equation}
    where $q(\gamma)=1/\gamma$. If $\varrho$ has the Lebesgue property, then
    \begin{equation}\label{eq:rock-ur}
        \widetilde{\varrho}_{\gamma}(\mathcal{X})=\inf_{x\in\mathbb{R}}\left(x+\gamma\varrho\left((\mathcal{X}-x)\vee0\right)\right)
    \end{equation}
    for each $\mathcal{X}\in L^{\infty}(\mathbb{P})$.
\end{proposition}
The representation (\ref{eq:rock-ur}) is a generalization of the Rockafellar-Uryasev \cite{rock-ur-general} formula for expected shortfall, a fundamental tool in the numerical estimation of expected shortfall.
\par
When $\varrho$ has the Lebesgue property, $\mathscr{Q}(\varrho)$ is $\sigma(L^{1},L^{\infty})$-compact (where we identify $\mathscr{Q}(\varrho)$ with the associated set of $\mathbb{P}$-densities), a consequence of the Jouini-Schachermayer-Touzi theorem (see \cite[Theorem 2.4]{owari-jfa}). Weak compactness of $\mathscr{Q}(\varrho)$ is used to prove (\ref{eq:rock-ur}) via a minimax argument (cf. \cite[Proposition 7]{partial-law-inv}).
\begin{proof}
    The representation (\ref{eq:es-rep-inflate}) is immediate from the fact that $\mathbb{S}\in\gamma\widetilde{\mathscr{Q}}(\varrho)\cap\mathscr{M}_{\mathbb{P}}$ iff there exists $\mathbb{Q}\in\mathscr{Q}(\varrho)$ with $\frac{d\mathbb{S}}{d\mathbb{Q}}\leq\gamma$, so we focus on (\ref{eq:rock-ur}).
    \par
    The Rockafellar-Uryasev representation of expected shortfall (see \cite{rock-ur-general}) gives
    \begin{equation*}
        \mathrm{ES}^{q(\gamma)}_{\mathbb{Q}}(\mathcal{X})=\inf_{x\in\mathbb{R}}\left(x+\frac{1}{q(\gamma)}\mathbb{E}^{\mathbb{Q}}\left((\mathcal{X}-x)\vee0\right)\right)=\inf_{x\in\mathbb{R}}\left(x+\gamma\mathbb{E}^{\mathbb{Q}}\left((\mathcal{X}-x)\vee0\right)\right)
    \end{equation*}
    for each $\mathbb{Q}\in\mathscr{M}_{\mathbb{P}}$. Hence, in light of (\ref{eq:es-rep-inflate}),
    \begin{equation*}
        \widetilde{\varrho}_{\gamma}(\mathcal{X})=\sup_{\mathbb{Q}\in\mathscr{Q}(\varrho)}\inf_{x\in\mathbb{R}}\Psi(\mathbb{Q},x)
    \end{equation*}
    where $\Psi(\mathbb{Q},x)=x+\gamma\mathbb{E}^{\mathbb{Q}}\left((\mathcal{X}-x)\vee0\right)$ for $(\mathbb{Q},x)\in\mathscr{Q}(\varrho)\times\mathbb{R}$, so the proof reduces to showing an appropriate minimax theorem applies to $\Psi$. For a fixed $x\in\mathbb{R}$, the map $\mathbb{Q}\longmapsto\Psi(\mathbb{Q},x)$ is affine and continuous when defined on the convex $\sigma(L^{1},L^{\infty})$-compact set $\mathscr{Q}(\varrho)$. For a fixed $\mathbb{Q}\in\mathscr{Q}(\varrho)$, $x\longmapsto\Psi(\mathbb{Q},x)$ is convex and continuous. These facts jointly imply Sion's minimax theorem is applicable (see Komiya \cite{ElementaryProofSions}), so that
    \begin{equation*}
        \widetilde{\varrho}_{\gamma}(\mathcal{X})=\inf_{x\in\mathbb{R}}\sup_{\mathbb{Q}\in\mathscr{Q}(\varrho)}\Psi(\mathbb{Q},x)=\inf_{x\in\mathbb{R}}\left(x+\gamma\varrho\left((\mathcal{X}-x)\vee0\right)\right),
    \end{equation*}
    as desired.
\end{proof}
\par
Before consideration of the risk sharing problem for inflations of risk measures, we briefly study the behavior of $\gamma$-inflation as a function of $\gamma$, which has instrumental significance.
\begin{proposition}\label{prop:inflate-continuous}
    Fix $\mathcal{X}\in L^{\infty}(\mathbb{P})$, and define $\varphi:[1,\infty)\longrightarrow\mathbb{R}$ by 
    \begin{equation*}
        \forall\gamma\in[1,\infty):\varphi(\gamma)=\widetilde{\varrho}_{\gamma}(\mathcal{X}).
    \end{equation*}
    Then $\varphi$ has the following properties:
    \begin{enumerate}
        \item $\varphi$ is non-decreasing.
        \item $\varphi$ is left continuous everywhere, and right continuous at $1$, in the sense that
        \begin{equation*}
            \lim_{\gamma\to1^{+}}\varphi(\gamma)=\varphi(1)=\varrho(\mathcal{X}).
        \end{equation*}
        \item If $\varrho$ has the Lebesgue property, then $\varphi$ is concave and continuous on $[1,\infty)$.
        \item As $\gamma\to\infty$, $\varphi(\gamma)\to\mathbb{P}-\mathrm{essup}\mathcal{X}$.
    \end{enumerate}
\end{proposition}
The interpretation of Proposition \ref{prop:inflate-continuous} is straightforward. Point (1) states that, as the uncertainty parameter increases, so does the estimated risk. Point (2) implies that, as the uncertainty $\gamma$ becomes close to the certain regime $\gamma=1$, the estimated risk approaches the estimated risk calculated without model uncertainty. Point (4) implies that, for very large values of the uncertainty parameter, the estimated risk will approach the worst possible risk.
\begin{proof}
    Since $\gamma\widetilde{\mathscr{Q}}(\varrho)\cap \mathscr{M}_{\mathbb{P}}\subseteq \gamma'\widetilde{\mathscr{Q}}(\varrho)\cap \mathscr{M}_{\mathbb{P}}$ whenever $\gamma\leq\gamma'$, it follows that $\varphi(\gamma)\leq\varphi(\gamma')$, showing (1).
    \par
    We now show (2), beginning with right continuity at $1$. By cash additivity, we may assume without loss of generality that $\mathcal{X}\geq0$. If $\mathbb{S}\in \gamma\widetilde{\mathscr{Q}}(\varrho)\cap \mathscr{M}_{\mathbb{P}}$, then there exists $\mathbb{Q}\in\mathscr{Q}(\varrho)$ so that $0\leq\frac{d\mathbb{S}}{d\mathbb{P}}\leq\gamma\frac{d\mathbb{Q}}{d\mathbb{P}}$. Hence, as $\mathcal{X}\geq0$,
    \begin{equation*}
        \mathbb{E}^{\mathbb{S}}(\mathcal{X})\leq\gamma\mathbb{E}^{\mathbb{Q}}(\mathcal{X})\leq\gamma\varrho(\mathcal{X}).
    \end{equation*}
    Taking the supremum over $\mathbb{S}\in \gamma\widetilde{\mathscr{Q}}(\varrho)\cap \mathscr{M}_{\mathbb{P}}$ and noting that $\varrho(\mathcal{X})\leq\varphi(\gamma)$, we obtain the inequality
    \begin{equation*}
        \varrho(\mathcal{X})\leq\varphi(\gamma)\leq\gamma\varrho(\mathcal{X}).
    \end{equation*}
    Taking $\gamma\downarrow1$ and applying the squeeze theorem yields right continuity at $1$.
    \par
    We now move on to left continuity. For each $\mathbb{Q}\in\mathscr{M}_{\mathbb{P}}$, the map $\gamma\longmapsto\mathrm{ES}^{q(\gamma)}_{\mathbb{Q}}(\mathcal{X})$ is continuous (a fact which follows easily from an alternate definition of expected shortfall; see \cite{es-dual}). From (\ref{eq:es-rep-inflate}), we therefore get that $\gamma\longmapsto\varphi(\gamma)$ is a supremum of continuous functions of $\gamma$. So $\varphi$ is lower semicontinuous and non-decreasing, implying left continuity.
    \par
    For (3), concavity of $\varphi$ on $[1,\infty)$ is immediate from (\ref{eq:rock-ur}). Hence $\varphi$ is continuous on $(1,\infty)$, and the only thing left to show is continuity at $1$, which was established in a previous part of the proposition.
    \par
    We now show (4). We have that
    \begin{equation}\label{eq:gamma-large-ineq}
        \mathrm{ES}^{1/\gamma}_{\mathbb{P}}\leq\varphi(\gamma)\leq\mathbb{P}-\mathrm{essup}\mathcal{X}
    \end{equation}
    where the inequality on the left side follows from the assumption $\mathbb{P}\in\mathscr{Q}(\varrho)$. Since $\lim_{\alpha\to0^{+}}\mathrm{ES}^{\alpha}_{\mathbb{P}}(\mathcal{X})=\mathbb{P}-\mathrm{essup}\mathcal{X}$, taking $\gamma\to\infty$ in (\ref{eq:gamma-large-ineq}) and applying the squeeze theorem yields (4).
\end{proof}
\subsubsection{Risk Sharing with Inflated Risk Measures}\label{subsubsec:inflate-share}
For a family $(\gamma_{a})_{a\in A}\in[1,\infty)^{A}$ and a coherent risk measure $\varrho$ it is possible to construct a collection $(\varrho_{a})_{a\in A}$ of risk preferences via $\varrho_{a}=\widetilde{\varrho}_{\gamma_{a}}$. Given such a family, we now consider the value function of the risk sharing problem, characterizing the dual representation of the value function, and ensuring $(\varrho_{a})_{a\in A}$ satisfies the requisite measurability condition under broad circumstances.
\begin{theorem}\label{thm:inflate-dual}
    Let $\varrho$ be a risk measure with the Lebesgue property and the representation (\ref{eq:dualsetrep}). Let $(\gamma_{a})_{a\in A}\in[1,\infty)^{A}$ be an $\mathscr{A}$-measurable map, with $\mu$-essential infimum $\Gamma$. Defining $\varrho_{a}=\widetilde{\varrho}_{\gamma_{a}}$ for each $a\in A$, we have the following.
    \begin{enumerate}
        \item The indexed collection $(\varrho_{a})_{a\in A}$ of risk measures is $\mathscr{A}$-measurable.
        \item The integral infimal convolution $\operatorname{\Box}_{a\in A}\varrho_{a}\mu(da)$ satisfies
        \begin{equation*}
            \operatorname{\Box}_{a\in A}\varrho_{a}\mu(da)=\widetilde{\varrho}_{\Gamma}.
        \end{equation*}
        In particular,
        \begin{equation*}
            \mathscr{Q}\left( \operatorname{\Box}_{a\in A}\varrho_{a}\mu(da)\right)=\Gamma\widetilde{\mathscr{Q}}(\varrho)\cap\mathscr{M}_{\mathbb{P}}.
        \end{equation*}
    \end{enumerate}
\end{theorem}
\begin{proof}
    For (1), we may find an increasing sequence $\left((\gamma^{n}_{a})_{a\in A}\right)_{n=1}^{\infty}$ of $\mathscr{A}$-measurable simple functions, such that $\gamma^{n}\uparrow\gamma$ pointwise. Furthermore, we may assume that $\gamma^{n}$ takes values in $[1,\infty)$ (indeed, one can replace $\gamma^{n}$ with the $\mathscr{A}$-measurable simple function $\gamma^{n}\vee1$). By the argument in Example \ref{ex:multiple-risks}, for each $n$, the family $(\varrho^{n}_{a})_{a\in A}$ of risk measures defined by $\varrho^{n}_{a}=\widetilde{\varrho}_{\gamma^{n}_{a}}$ for each $a\in A$ is $\mathscr{A}$-measurable. Thus, for each $\mathscr{A}$-measurable allocation $(X_{a})_{a\in A}$, $a\longmapsto\varrho^{n}_{a}(X_{a})$ is $\mathscr{A}$-measurable. As $n\to\infty$, Proposition \ref{prop:inflate-continuous} implies that $\lim_{n\to\infty}\varrho^{n}_{a}(X_{a})$ exists and equals $\varrho_{a}(X_{a})$. Since pointwise limits of $\mathscr{A}$-measurable functions are $\mathscr{A}$-measurable, this implies that $a\longmapsto\varrho_{a}(X_{a})$ is $\mathscr{A}$-measurable. Since $(X_{a})_{a\in A}$ was an arbitrary $\mathscr{A}$-measurable allocation, this shows that $(\varrho_{a})_{a\in A}$ is an $\mathscr{A}$-measurable collection of risk measures, proving (1).
    \item
    To prove (2), we apply Theorem \ref{thm:dual-rep-represent}. First, one must show that the preconditions for Theorem \ref{thm:dual-rep-represent} hold. Thus, one must establish the following:
    \begin{enumerate}[label=\upshape\roman*.,ref=\upshape\roman*]
        \item\label{item:int-finite} $\int_{A}\vert{\varrho_{a}}(0)\vert\mu(da)<\infty$.
        \item\label{item:lebesgue} The collection $(\varrho_{a})_{a\in A}$ consists of risk measures with the Lebesgue property.
        \item\label{item:glob-finite} The integral infimal convolution $\operatorname{\Box}_{a\in A}\varrho_{a}\mu(da)$ is globally finite.
    \end{enumerate}
    Clearly, since $\varrho_{a}(0)=0$ for all $a$, (\ref*{item:int-finite}) holds. For (\ref*{item:lebesgue}), note that the Jouini-Schachermayer-Touzi theorem (see \cite[Theorem 2.4]{owari-jfa}) implies that, since $\varrho$ has the Lebesgue property, $\mathscr{Q}(\varrho)$ must be uniformly integrable (viewed as a subset of $L^{1}(\mathbb{P})$ via the Radon-Nikodým derivative). Thus, for each $\gamma'\geq1$, $\gamma'\widetilde{\mathscr{Q}}(\varrho)\cap\mathscr{M}_{\mathbb{P}}$ is uniformly integrable. Since any risk measure representable as a supremum of expectations over a uniformly integrable set of probability measures has the Lebesgue property, it follows that every inflation of $\varrho$ has the Lebesgue property. In particular, $(\varrho_{a})_{a\in A}$ consists of risk measures with the Lebesgue property, and (\ref*{item:lebesgue}) therefore holds. To establish (\ref*{item:glob-finite}), it suffices to verify the preconditions of Proposition \ref{prop:glob-finite}; (\ref*{item:int-finite}) and (\ref*{item:lebesgue}) are both preconditions (both of which we have already verified), and the only remaining precondition is the existence of $\mathbb{Q}\in\bigcap_{a\in A}\{\varrho^{\ast}_{a}<\infty\}$ with $\int_{A}\varrho^{\ast}_{a}(\mathbb{Q})\mu(da)<\infty$. For this last precondition, fix some $\mathbb{Q}\in\{\varrho^{\ast}<\infty\}\neq\emptyset$. For each $a\in A$, $\varrho^{\ast}_{a}(\mathbb{Q})=0$, proving the claim.
    \par
    We now apply Theorem \ref{thm:dual-rep-represent}. It suffices to show that $\mathbb{Q}\in\Gamma\widetilde{\mathscr{Q}}(\varrho)\cap\mathscr{M}_{\mathbb{P}}$ if, and only if, $\left(\operatorname{\Box}_{a\in A}\varrho_{a}\mu(da)\right)^{\ast}(\mathbb{Q})<\infty$. If $\mathbb{Q}\in\Gamma\widetilde{\mathscr{Q}}(\varrho)\cap\mathscr{M}_{\mathbb{P}}$, then $\varrho^{\ast}_{a}(\mathbb{Q})=0$ for $\mu$-a.e. $a\in A$, implying (via Theorem \ref{thm:dual-rep-represent}) that $\left(\operatorname{\Box}_{a\in A}\varrho_{a}\mu(da)\right)^{\ast}(\mathbb{Q})<\infty$. Conversely, suppose $\left(\operatorname{\Box}_{a\in A}\varrho_{a}\mu(da)\right)^{\ast}(\mathbb{Q})<\infty$. It is easy to see that $\mathbb{Q}\in\gamma\widetilde{\mathscr{Q}}(\varrho)\cap\mathscr{M}_{\mathbb{P}}$ for each $\gamma>\Gamma$. Thus,
    \begin{equation*}
        \mathbb{Q}\in\bigcap_{\gamma>\Gamma}\gamma\widetilde{\mathscr{Q}}(\varrho)\cap\mathscr{M}_{\mathbb{P}}=\mathscr{M}_{\mathbb{P}}\cap\bigcap_{\gamma>\Gamma}\gamma\widetilde{\mathscr{Q}}(\varrho),
    \end{equation*}
    implying it suffices to show that $\Gamma\widetilde{\mathscr{Q}}(\varrho)=\bigcap_{\gamma>\Gamma}\gamma\widetilde{\mathscr{Q}}(\varrho)$. Fix any $\mathcal{X}\in\bigcap_{\gamma>\Gamma}\gamma\widetilde{\mathscr{Q}}(\varrho)$; it suffices to show that $\mathcal{X}\in\Gamma\widetilde{\mathscr{Q}}(\varrho)$. Take a strictly decreasing $(\Gamma_{n})_{n=1}^{\infty}\downarrow\Gamma$; for each $n$, we may find $\mathbb{Q}_{n}\in\mathscr{Q}(\varrho)$ with
    \begin{equation*}
        0\leq\mathcal{X}\leq\Gamma_{n}\frac{d\mathbb{Q}_{n}}{d\mathbb{P}}.
    \end{equation*}
    As established before, $\mathscr{Q}(\varrho)$ is uniformly integrable. Thus, by the Dunford-Pettis theorem, after passing to subsequences, we may assume $(\mathbb{Q}_{n})_{n=1}^{\infty}$ converges to some $\mathbb{Q}\in\mathscr{M}_{\mathbb{P}}$ in $\sigma(L^{1},L^{\infty})$; since $\mathscr{Q}(\varrho)$ is weakly closed in $L^{1}(\mathbb{P})$, $\mathbb{Q}\in\mathscr{Q}(\varrho)$. As the positive cone in $L^{1}(\mathbb{P})$ is weakly closed,
    \begin{equation*}
        0\leq\mathcal{X}\leq\Gamma\frac{d\mathbb{Q}}{d\mathbb{P}},
    \end{equation*}
    showing that $\mathcal{X}\in\Gamma\widetilde{\mathscr{Q}}(\varrho)$, as desired.
\end{proof}
We consider now whether the infimum inherent in the value function is attained, in the setting of Theorem \ref{thm:inflate-dual}. There are two circumstances to consider, depending on the nature of the essential infimum $\Gamma$ of $(\gamma_{a})_{a\in A}$:
\begin{enumerate}
    \item $\mu(\{a:\gamma_{a}=\Gamma\})>0$, in which case an optimal allocation is found by giving all the risk to the agents $a$ such that $\gamma_{a}=\Gamma$.
    \item $\mu(\{a:\gamma_{a}=\Gamma\})=0$, in which case the existence of optimal allocations becomes subtle. Intuitively, the infimum should not be attained, since one should be able to shift risk from agents $a$ with $\gamma_{a}>\Gamma+\varepsilon$ (where $0<\varepsilon\ll1$) to agents $b$ with $\gamma_{b}\leq\Gamma+\varepsilon$, constituting an improvement on an apparently optimal allocation (this intuition is formalized in Appendix \ref{sec:prf-attain-notattain}). However, the infimum is always attained if the risk $\mathcal{X}$ to be allocated is a constant random variable, or more generally if $\gamma\longmapsto\widetilde{\varrho}_{\gamma}(\mathcal{X})$ is constant on $[\Gamma,\infty)$. Thus, to conclude an optimal allocation does not exist, one must introduce a condition on $\mathcal{X}$ ensuring it is not unaffected by a change in the parameter $\gamma$.
\end{enumerate}
Our main result in this direction is Theorem \ref{thm:attain-notattain} below. Compared to the finite agent case for expected shortfall (see \cite{quantile-share}), our result simultaneously exhibits new phenomena and generalizes known results: when $\mu(\{a:\gamma_{a}=\Gamma\})>0$, the finite agent formulas for an optimal allocation remain true, while if $\mu(\{a:\gamma_{a}=\Gamma\})=0$, an optimal allocation may fail to exist, something which is not true in the finite agent case.
\begin{theorem}\label{thm:attain-notattain}
    Let $\varrho$ be a risk measure with the Lebesgue property and the representation (\ref{eq:dualsetrep}). Let $(\gamma_{a})_{a\in A}\in[1,\infty)^{A}$ be an $\mathscr{A}$-measurable map, with $\mu$-essential infimum $\Gamma$. Defining $\varrho_{a}=\widetilde{\varrho}_{\gamma_{a}}$ for each $a\in A$, we have the following.
    \begin{enumerate}
        \item Suppose $\mu(\{a:\gamma_{a}=\Gamma\})>0$. Then, for any $\mathcal{X}\in L^{\infty}(\mathbb{P})$, the allocation
        \begin{equation*}
            (\mathbf{1}_{\{b:\gamma_{b}=\Gamma\}}(a)\mathcal{X}/\mu(\{b:\gamma_{b}=\Gamma\}))_{a\in A}\in\mathbb{A}(\mathcal{X})
        \end{equation*}
        is optimal, in the sense that
        \begin{equation*}
            \left(\operatorname{\Box}_{a\in A}\varrho_{a}\mu(da)\right)(\mathcal{X})=\int_{A}\varrho_{a}(\mathbf{1}_{\{b:\gamma_{b}=\Gamma\}}(a)\mathcal{X}/\mu(\{b:\gamma_{b}=\Gamma\}))\mu(da).
        \end{equation*}
        \item Suppose the following conditions are true for $\mathcal{X}\in L^{\infty}(\mathbb{P})$.
        \begin{enumerate}
            \item $\mu(\{a:\gamma_{a}=\Gamma\})=0$.
            \item There exists $\Gamma'>\Gamma$ so that $\widetilde{\varrho}_{\Gamma}(\mathcal{X})\neq\widetilde{\varrho}_{\Gamma'}(\mathcal{X})$.
        \end{enumerate}
        Then there does not exist an allocation $(X_{a})_{a\in A}\in\mathbb{A}(\mathcal{X})$ such that
        \begin{equation*}
            \left(\operatorname{\Box}_{a\in A}\varrho_{a}\mu(da)\right)(\mathcal{X})=\int_{A}\varrho_{a}(X_{a})\mu(da).
        \end{equation*}
    \end{enumerate}
\end{theorem}
\begin{proof}
    We prove the first assertion here; the proof of the second assertion is contained in Appendix \ref{sec:prf-attain-notattain}.
    \par
    For the first assertion, note that Theorem \ref{thm:inflate-dual} implies equivalence to the claim that
    \begin{equation*}
        \widetilde{\varrho}_{\Gamma}(\mathcal{X})=\int_{A}\varrho_{a}(\mathbf{1}_{\{b:\gamma_{b}=\Gamma\}}(a)\mathcal{X}/\mu(\{b:\gamma_{b}=\Gamma\}))\mu(da).
    \end{equation*}
    Clearly,
    \begin{equation*}
        \int_{A}\varrho_{a}(\mathbf{1}_{\{b:\gamma_{b}=\Gamma\}}(a)\mathcal{X}/\mu(\{b:\gamma_{b}=\Gamma\}))\mu(da)=\int_{\{b:\gamma_{b}=\Gamma\}}\frac{1}{\mu(\{b:\gamma_{b}=\Gamma\})}\widetilde{\varrho}_{\Gamma}\left(\mathcal{X}\right)\mu(da)
    \end{equation*}
    \begin{equation*}
        =\widetilde{\varrho}_{\Gamma}(\mathcal{X}),
    \end{equation*}
    as desired, proving the first assertion.
\end{proof}
The non-constancy precondition for non-existence in Theorem \ref{thm:attain-notattain}, which asks that $\widetilde{\varrho}_{\Gamma}(\mathcal{X})\neq\widetilde{\varrho}_{\Gamma'}(\mathcal{X})$ for some $\Gamma'>\Gamma$, is always satisfied when the law of $\mathcal{X}$ is non-atomic (see Proposition \ref{prop:essup-and-inflation}). Thus, in most situations of interest, the aforementioned constancy subtlety is immaterial, and non-existence of optimizers holds whenever $\mu(\{a:\gamma_{a}=\Gamma\})=0$.
\section{Applications to Pareto Efficiency}\label{sec:pareto}
In this section, we exhibit an economic interpretation of the integral infimal convolution: $(X_{a})_{a\in A}$ is an optimizer of the integral infimal convolution iff $(X_{a})_{a\in A}$ is Pareto efficient. The non-existence results of Theorem \ref{thm:attain-notattain} therefore show many markets with Pareto optima in the finite agent setting—such as those based on expected shortfall—do not admit Pareto optima if there are a continuum of agents. This is in line with the economics literature, which has shown that the set of Pareto optima shrinks as the number of agents increases (see \cite{aumanncontinuum,limit-debreu-scarf}).
\par
We first recall some terminology from economic theory (see \cite[\S1.2]{econ-analysis-law}).
\begin{definition}\label{def:pareto}
    Let $(\varrho_{a})_{a\in A}$ be an $\mathscr{A}$-measurable collection of preferences. An $\mathcal{X}$-feasible allocation $(X_{a})_{a\in A}$ is said to be Pareto efficient if there does not exist another $\mathcal{X}$-feasible allocation $(Y_{a})_{a\in A}$ with $\varrho_{a}(Y_{a})\leq\varrho_{a}(X_{a})$ $\mu$-a.e. and $\mu\left(\left\{\varrho_{a}(Y_{a})<\varrho_{a}(X_{a})\right\}\right)>0$. If such a $(Y_{a})_{a\in A}$ exists for an $\mathcal{X}$-feasible $(X_{a})_{a\in A}$, $(Y_{a})_{a\in A}$ is said to be a Pareto improvement on $(X_{a})_{a\in A}$.
\end{definition}
Definition \ref{def:pareto}, after admitting the necessary refinements to adjust for $\mu$-null sets, is conceptually the same as Pareto efficiency in the finite agent setting. Similar measure-based definitions have appeared before in the economics literature (see \cite[Definition 5.4]{sagara-variational}).
\par
It is known that Pareto efficiency in the finite agent case is related to the optimization of a social welfare function via the Negishi weight method, and that, under translation invariance of the preferences, this reduces to the classical risk sharing problem (see \cite[Proposition 3.3]{ghossoubzhu2025}). Similarly, we have the following result, essentially due to Ghossoub and Nendel \cite{gn-pareto}, which gives an economic import to the integral infimal convolution.
\begin{proposition}\label{prop:pareto-characterize}
    Suppose $(\varrho_{a})_{a\in A}$ is an $\mathscr{A}$-measurable collection of risk measures, $\mathcal{X}\in L^{\infty}(\mathbb{P})$, and
    \begin{equation*}
        -\infty<\left(\operatorname{\Box}_{a\in A}\varrho_{a}\mu(da)\right)(\mathcal{X})<\infty.
    \end{equation*}
    Then an $\mathcal{X}$-feasible allocation $(X_{a})_{a\in A}$ satisfying $\int_{A}\min\left\{0,\varrho_{a}(X_{a})\right\}\mu(da)>-\infty$ is Pareto efficient if, and only if,
    \begin{equation*}
        \left(\operatorname{\Box}_{a\in A}\varrho_{a}\mu(da)\right)(\mathcal{X})=\int_{A}\varrho_{a}(X_{a})\mu(da).
    \end{equation*}
\end{proposition}
The condition that $\int_{A}\min\left\{0,\varrho_{a}(X_{a})\right\}\mu(da)>-\infty$ is minor. In fact, it is satisfied for every $\mathcal{X}$-feasible allocation if there exists $\xi\in L^{1}(\mu)$ and $\mathbb{Q}\in\mathscr{M}_{\mathbb{P}}$ such that $\varrho^{\ast}_{a}(\mathbb{Q})\leq\xi(a)$ for $\mu$-a.e. $a\in A$ (cf. Proposition \ref{prop:glob-finite}). So in essentially all situations of interest, Proposition \ref{prop:pareto-characterize} provides a complete characterization of Pareto optima.
\par
One direction of Proposition \ref{prop:pareto-characterize} is easy: optimizers of the integral infimal convolution, just as in the finite agent setting, are always Pareto optimal. The converse is more subtle and relies on cash additivity of preferences.
\par
In the finite agent setting, if cash additivity is not assumed, the analogue of Proposition \ref{prop:pareto-characterize} requires a hyperplane separation argument (see \cite[Proposition 2.5]{ghossoubzhu2025}). If neither cash additivity nor convexity is assumed, there is no analogue to Proposition \ref{prop:pareto-characterize} (see \cite[pp. 558-561]{micro-theory}).
\begin{proof}
    If $\left(\operatorname{\Box}_{a\in A}\varrho_{a}\mu(da)\right)(\mathcal{X})=\int_{A}\varrho_{a}(X_{a})\mu(da)$, and there exists a Pareto improvement $(Y_{a})_{a\in A}$ on $(X_{a})_{a\in A}$, then
    \begin{equation*}
        \left(\operatorname{\Box}_{a\in A}\varrho_{a}\mu(da)\right)(\mathcal{X})\leq\int_{A}\varrho_{a}(Y_{a})\mu(da)<\int_{A}\varrho_{a}(X_{a})\mu(da)=\left(\operatorname{\Box}_{a\in A}\varrho_{a}\mu(da)\right)(\mathcal{X}),
    \end{equation*}
    a contradiction.
    \par
    Conversely, suppose $(X_{a})_{a\in A}$ is Pareto efficient, but
    \begin{equation*}
        \left(\operatorname{\Box}_{a\in A}\varrho_{a}\mu(da)\right)(\mathcal{X})<\int_{A}\varrho_{a}(X_{a})\mu(da).
    \end{equation*}
    Hence, there exists an $\mathcal{X}$-feasible $(Y_{a})_{a\in A}$ with
    \begin{equation*}
        \int_{A}\varrho_{a}(Y_{a})\mu(da)<\int_{A}\varrho_{a}(X_{a})\mu(da).
    \end{equation*}
    Because $\int_{A}\min\left\{0,\varrho_{a}(X_{a})\right\}\mu(da)>-\infty$, there exists $m\in\mathbb{R}$ so that
    \begin{equation*}
        \int_{A}\varrho_{a}(Y_{a})\mu(da)<\int_{A}\min\left\{m,\varrho_{a}(X_{a})\right\}\mu(da).
    \end{equation*}
    Define $R=\frac{1}{\mu(A)}\int_{A}\left(\min\left\{m,\varrho_{a}(X_{a})\right\}-\varrho_{a}(Y_{a})\right)\mu(da)>0$.
    Consider $(Z_{a})_{a\in A}\in\mathbb{A}(\mathcal{X})$ defined as
    \begin{equation*}
        Z_{a}=Y_{a}+\min\left\{m,\varrho_{a}(X_{a})\right\}-\varrho_{a}(Y_{a})-R.
    \end{equation*}
    For each $a\in A$,
    \begin{equation*}
        \varrho_{a}(Z_{a})+R=\varrho_{a}\left(Y_{a}+\min\left\{m,\varrho_{a}(X_{a})\right\}-\varrho_{a}(Y_{a})\right)=\min\left\{m,\varrho_{a}(X_{a})\right\}\leq\varrho_{a}(X_{a}).
    \end{equation*}
    Hence, as $R>0$, for each $a\in A$, $\varrho_{a}(Z_{a})\leq\varrho_{a}(X_{a})-R<\varrho_{a}(X_{a})$, showing $(Z_{a})_{a\in A}$ constitutes a Pareto improvement on $(X_{a})_{a\in A}$, contradicting the Pareto efficiency of $(X_{a})_{a\in A}$.
\end{proof}
\begin{remark}
    The second part of the above argument is based on Kaldor-Hicks efficiency (for a definition and a discussion of Kaldor-Hicks efficiency, see \cite[\S1.2]{econ-analysis-law}). More precisely, it shows that if $(Y_{a})_{a\in A}\in\mathbb{A}(\mathcal{X})$ has a lower total risk than $(X_{a})_{a\in A}\in\mathbb{A}(\mathcal{X})$, it is possible for the beneficiaries of the lower risk total from $(Y_{a})_{a\in A}$ to compensate with cash those that are left worse off from $(Y_{a})_{a\in A}$.
    \qed
\end{remark}
Conjoining Proposition \ref{prop:pareto-characterize} with the ill-posedness results of \S\ref{sec:examples}, we obtain a non-existence result for Pareto optima.
\begin{theorem}\label{thm:es-no-optima}
    Suppose $(\beta_{a})_{a\in A}\in(0,1]^{A}$ is an $\mathscr{A}$-measurable map with $\mu$-essential supremum $\beta>0$, and $\mathcal{X}\in L^{\infty}(\mathbb{P})$ is an aggregate endowment. Define the individual preferences $(\varrho_{a})_{a\in A}$ by $\varrho_{a}=\mathrm{ES}^{\beta_{a}}_{\mathbb{P}}$. If $\mu(\{a:\beta_{a}=\beta\})=0$ and there exists $0<\alpha<\beta$ so $\mathrm{ES}^{\beta}_{\mathbb{P}}(\mathcal{X})\neq\mathrm{ES}^{\alpha}_{\mathbb{P}}(\mathcal{X})$, then there does not exist any Pareto efficient $\mathcal{X}$-feasible allocations.
\end{theorem}
\begin{proof}
    Theorem \ref{thm:attain-notattain} implies there cannot exist $(X_{a})_{a\in A}\in\mathbb{A}(\mathcal{X})$ with
    \begin{equation}\label{eq:is-maximizer}
        \int_{A}\varrho_{a}(X_{a})\mu(da)=\left(\operatorname{\Box}_{a\in A}\varrho_{a}\mu(da)\right)(\mathcal{X}).
    \end{equation}
    Since any Pareto efficient allocation of $\mathcal{X}$ must satisfy (\ref{eq:is-maximizer}) (see Proposition \ref{prop:pareto-characterize}), it follows that Pareto efficient allocations of $\mathcal{X}$ cannot exist.
\end{proof}
\begin{example}\label{ex:no-pareto-es}
    Consider the agent space $A=\left[\underline{\alpha},\overline{\alpha}\right]$, with $\mu$ the uniform measure on this interval, where $\underline{\alpha}<\overline{\alpha}$ and $\underline{\alpha},\overline{\alpha}\in(0,1]$ (this is essentially the setup of Example \ref{ex:aumman}). Suppose the aggregate endowment $\mathcal{X}$ is uniformly distributed on $[0,1]$, and agent preferences $(\varrho_{a})_{a\in A}$ are given by $\varrho_{a}=\mathrm{ES}^{a}_{\mathbb{P}}$. Then, by Theorem \ref{thm:es-no-optima}, no Pareto efficient allocations of $\mathcal{X}$ exist.
    \qed
\end{example}
In contrast to the above example, Pareto optima always exist for expected shortfall preferences if the agent set is finite (see \cite{quantile-share}).
\appendix
\section{The Weak-Star Strassen Theorem}\label{sec:tech}
Theorem 2.2 of Hiai and Umegaki \cite{hiai-umegaki} establishes an integral exchange formula for correspondences valued in a separable Banach space, allowing one to swap an infimum (equivalently, a supremum) and an integral, allowing one to characterize the support function of certain set integrals. The Hiai-Umegaki result is an example of a Strassen-type theorem. Strassen-type theorems have been extended to cover correspondences valued in separable Banach spaces equipped with the weak topology (see \cite{strass-pettis}), correspondences valued in the compact subsets of a locally convex topological vector space (see \cite{strass-comp}), and have ramifications even in the finite dimensional case (see \cite{aumann-exchange}). However, the literature on this topic is both highly technical and likely not directly applicable to our circumstances (e.g., requiring the correspondence to take weak-star compact values). Thus, in this section, we derive a Strassen-type theorem for correspondences valued in the dual of a separable Banach space and taking weak-star closed values, with measurability and integration understood in a weak-star sense.
\subsection{Notation}
Let $(A,\mathscr{A},\mu)$ denote a finite complete measure space. The trace $\sigma$-algebra $\{C\cap B:C\in\mathscr{A}\}$ of $B\in\mathscr{A}$ is denoted $\mathscr{A}_{B}$. Let $E$ be a separable Banach space; denote by $B_{E}$ the unit ball of $E$, and let $B^{\ast}=B^{\circ}_{E}$ denote the closed unit ball of $E^{\ast}$, the dual of $E$.
\par
A function $f:A\longrightarrow E^{\ast}$ is said to be $\mathscr{A}$-measurable if $a\longmapsto\langle{x,f(a)}\rangle$ is $\mathscr{A}$-measurable for each $x\in E$. An $\mathscr{A}$-measurable function $f:A\longrightarrow E^{\ast}$ is said to be Gelfand integrable if $\langle{x,f}\rangle\in L^{1}(\mu)$ for each $x\in E$. If $f$ is Gelfand integrable, and $B\in\mathscr{A}$, there exists a unique element $g_{B}\in E^{\ast}$ such that $\langle{x,g_{B}}\rangle=\int_{B}\langle{x,f(a)}\rangle\mu(da)$. The element $g_{B}$ is denoted $\int_{B}f(a)\mu(da)$, and is called the Gelfand integral of $f$ over $B$. These notions all parallel those introduced for $E^{\ast}=L^{\infty}(\mathbb{P})$ in \S\ref{subsec:alloc}.
\par
Consider a correspondence $F:A\longrightarrow 2^{E^{\ast}}$. Given a subset $U\subseteq E^{\ast}$, define
\begin{equation*}
    F^{-1}(U)=\{a\in A:F(a)\cap U\neq\emptyset\}.
\end{equation*}
An integrable selector of $F$ is a Gelfand integrable function $f:A\longrightarrow E^{\ast}$ such that $f(a)\in F(a)$ for $\mu$-a.e. $a\in A$. The set of integrable selectors of $F$ is denoted $S^{1}(F)$.
\begin{definition}\label{def:aumann-int}
    The Aumann integral of $F$, denoted $\int_{A}F(a)\mu(da)$, is the subset of $E^{\ast}$ defined by
    \begin{equation*}
        \int_{A}F(a)\mu(da)=\left\{\int_{A}f(a)\mu(da):f\in S^{1}(F)\right\}.
    \end{equation*}
\end{definition}
In a similar fashion to the above concept of integration, one can introduce notions of measurability for correspondences (see \cite[p. 592 \textit{et seq}]{aliprantis-inf-dim}).
\begin{definition}\label{def:corr-measurable}
    $F$ is said to be $\mathscr{A}$-measurable if $F^{-1}(U)\in\mathscr{A}$ for every $\sigma(E^{\ast},E)$-closed $U\subseteq E^{\ast}$.
\end{definition}
\begin{definition}\label{def:effros}
    $F$ is said to be Effros $\mathscr{A}$-measurable if $F^{-1}(U)\in\mathscr{A}$ for every $\sigma(E^{\ast},E)$-open $U\subseteq E^{\ast}$.
\end{definition}
Definition \ref{def:corr-measurable} and Definition \ref{def:effros} are coherent for correspondences valued in any topological space, not just $(E^{\ast},\sigma(E^{\ast},E))$. In the sequel, we generally employ $\mathscr{A}$-measurability rather than Effros $\mathscr{A}$-measurability, since the former has better stability properties. However, since many results are stated in terms of Effros $\mathscr{A}$-measurability, we cannot expunge Definition \ref{def:effros} from our analysis.
\subsection{Preliminary Results}
In this subsection, we use $F$ to denote an $\mathscr{A}$-measurable correspondence $A\longrightarrow 2^{E^{\ast}}$ with non-empty $\sigma(E^{\ast},E)$-closed values.
\par
For $\lambda\geq0$, define a correspondence $F_{\lambda}:A\longrightarrow 2^{\lambda B^{\ast}}$ by $F_{\lambda}=F\cap\lambda B^{\ast}$. $F_{\lambda}$ is $\mathscr{A}$-measurable. Define $\widetilde{F}_{\lambda}$ as the restriction of $F_{\lambda}$ to $R_{\lambda}=F^{-1}(\lambda B^{\ast})\in\mathscr{A}$; it is easy to see that $\widetilde{F}_{\lambda}$ is $\mathscr{A}_{R_{\lambda}}$-measurable.
\begin{lemma}\label{lem:count-represent}
    There exists a collection $\{f_{n}:n\in\mathbb{N}\}\subseteq (E^{\ast})^{A}$ of $\mathscr{A}$-measurable functions such that
    \begin{equation*}
        F(a)=\overline{\bigcup_{n\in\mathbb{N}}\{f_{n}(a)\}}^{\sigma(E^{\ast},E)}
    \end{equation*}
    for each $a\in A$.
\end{lemma}
\begin{proof}
    Take $\lambda$ large enough so $R_{\lambda}\neq\emptyset$. By a result of Himmelberg \cite{himm-meas}, since $\lambda B^{\ast}$ is a Polish space and $\widetilde{F}_{\lambda}$ is $\mathscr{A}_{R_{\lambda}}$-measurable, there exists a collection $\{g_{n}^{\lambda}:n\in\mathbb{N}\}\subseteq (E^{\ast})^{R_{\lambda}}$ of $\mathscr{A}_{R_{\lambda}}$-measurable functions such that
    \begin{equation*}
        \widetilde{F}_{\lambda}(a)=\overline{\bigcup_{n\in\mathbb{N}}\left\{g_{n}^{\lambda}(a)\right\}}^{\sigma(E^{\ast},E)}
    \end{equation*}
    for each $a\in R_{\lambda}$. If there exists an $\mathscr{A}$-measurable $h:A\longrightarrow E^{\ast}$ such that $h(a)\in F(a)$ for each $a\in A$, the claim is proved. Indeed, take $(\lambda_{n})_{n=1}^{\infty}\uparrow\infty$, and consider the countable collection $\left\{k^{n}_{m}:(n,m)\in\mathbb{N}\times\mathbb{N}\right\}$ of $\mathscr{A}$-measurable functions defined by
    \begin{equation*}
        k^{n}_{m}|_{R_{\lambda_{n}}}=g^{\lambda_{n}}_{m},
    \end{equation*}
    \begin{equation*}
        k^{n}_{m}|_{A\setminus R_{\lambda_{n}}}=h.
    \end{equation*}
    Then $F(a)=\overline{\bigcup_{(n,m)\in\mathbb{N}\times\mathbb{N}}\{k^{n}_{m}(a)\}}^{\sigma(E^{\ast},E)}$, proving the claim, since one can consider a bijection $\mathbb{N}\longrightarrow\mathbb{N}\times\mathbb{N}$.
    \par
    We now prove the existence of an $\mathscr{A}$-measurable $h:A\longrightarrow E^{\ast}$ such that $h(a)\in F(a)$ for each $a\in A$. There exists a disjoint partition $\{D_{n}:n\in G\subseteq\mathbb{N}\}\subseteq\mathscr{A}\setminus\{\emptyset\}$ of $A$ such that for each $n\in G$, $D_{n}\subseteq F^{-1}(\lambda' B^{\ast})$ for large enough $\lambda'$ (which may depend on $n$).\footnote{For example, take $\widetilde{\lambda}$ large enough so that $R_{\widetilde{\lambda}}\neq\emptyset$. Define $D_{1}=F^{-1}\left(\widetilde{\lambda}B^{\ast}\right)$, and let $D_{n+1}=F^{-1}\left((n+1)\widetilde{\lambda}B^{\ast}\right)\setminus F^{-1}\left(n\widetilde{\lambda}B^{\ast}\right)$. Taking $G=\{n:D_{n}\neq\emptyset\}$ yields the desired construction.} Fix $n\in G$. Since $\lambda'B^{\ast}$ is a Polish space and $\widetilde{F}_{\lambda'}$ is $\mathscr{A}_{R_{\lambda'}}$-measurable (hence also Effros $\mathscr{A}_{R_{\lambda'}}$-measurable, see \cite[Lemma 18.2]{aliprantis-inf-dim}), the Kuratowski-Ryll-Nardzewski selection theorem (see \cite[p. 600]{aliprantis-inf-dim}) implies the existence of an $\mathscr{A}$-measurable $h_{n}:D_{n}\longrightarrow E^{\ast}$ with $h_{n}(a)\in F(a)$ for each $a\in D_{n}$. Allowing $n$ to vary, we may define $h$ by setting $h|_{D_{n}}=h_{n}$ for each $n\in G$.
\end{proof}
\begin{lemma}\label{lem:sup-measurable}
    Let $F$ be $\mathscr{A}$-measurable. Then, for every $x\in E$, the function
    \begin{equation*}
        a\longmapsto\sup_{x^{\ast}\in F(a)}\langle{x,x^{\ast}}\rangle
    \end{equation*}
    is $\mathscr{A}$-measurable.
\end{lemma}
\begin{proof}
    Let $\{f_{n}:n\in\mathbb{N}\}$ be as in Lemma \ref{lem:count-represent}. Since the map $x^{\ast}\longmapsto\langle{x,x^{\ast}}\rangle$ is $\sigma(E^{\ast},E)$-continuous for each $x\in E$, it follows that, for each $x\in E$,
    \begin{equation*}
        \sup_{x^{\ast}\in F(a)}\langle{x,x^{\ast}}\rangle=\sup_{n\in\mathbb{N}}\langle{x,f_{n}(a)}\rangle,
    \end{equation*}
    representing $a\longmapsto\sup_{x^{\ast}\in F(a)}\langle{x,x^{\ast}}\rangle$ as a countable supremum of $\mathscr{A}$-measurable functions.
\end{proof}
\subsection{Statement of the Result}
\begin{theorem}\label{thm:hiai-umegaki}
    Let $F$ be an $\mathscr{A}$-measurable correspondence with non-empty $\sigma(E^{\ast},E)$-closed values. Suppose $S^{1}(F)\neq\emptyset$. For all $x\in E$, we have that
    \begin{equation*}
        \sup_{x^{\ast}\in\int_{A}F(a)\mu(da)}\langle{x,x^{\ast}}\rangle=\int_{A}\sup_{x^{\ast}\in F(a)}\langle{x,x^{\ast}}\rangle\mu(da).
    \end{equation*}
\end{theorem}
Note that Lemma \ref{lem:sup-measurable} implies that $a\longmapsto\sup_{x^{\ast}\in F(a)}\langle{x,x^{\ast}}\rangle$ is $\mathscr{A}$-measurable, making the integral in Theorem \ref{thm:hiai-umegaki} above well-defined.
\par
Let us state a corollary to Theorem \ref{thm:hiai-umegaki}, which is more directly applicable to our situation with risk measures than Theorem \ref{thm:hiai-umegaki}. Rather than being interested in the supremum of $\langle{x,x^{\ast}}\rangle$ over $x^{\ast}\in\int_{A}F(a)\mu(da)$, we are interested in the supremum of a larger set $C\supseteq\int_{A}F(a)\mu(da)$.
\begin{theorem}\label{thm:hiai-umegaki-dense}
    Let $F$ be an $\mathscr{A}$-measurable correspondence with non-empty $\sigma(E^{\ast},E)$-closed values. Suppose $S^{1}(F)\neq\emptyset$, and let $C\subseteq E^{\ast}$ be such that $C=\overline{\int_{A}F(a)\mu(da)}^{\mathscr{T}}$ for some topology $\mathscr{T}$ finer than $\sigma(E^{\ast},E)$. For all $x\in E$, we have that
    \begin{equation*}
        \sup_{x^{\ast}\in C}\langle{x,x^{\ast}}\rangle=\int_{A}\sup_{x^{\ast}\in F(a)}\langle{x,x^{\ast}}\rangle\mu(da).
    \end{equation*}
\end{theorem}
\begin{proof}
    The claim is a trivial joint consequence of Theorem \ref{thm:hiai-umegaki} and $\mathscr{T}$-continuity of the map $x^{\ast}\longmapsto\langle{x,x^{\ast}}\rangle$.
\end{proof}
\subsection{Proof of Theorem \ref{thm:hiai-umegaki}}
\begin{proof}[Proof of Theorem \ref{thm:hiai-umegaki}]
Let $\Xi(a)=\sup_{x^{\ast}\in F(a)}\langle{x,x^{\ast}}\rangle$. If the claim were false, there would exist $\beta<\int_{A}\Xi(a)\mu(da)$ such that $\beta>\left\langle{x,\int_{A}g(a)\mu(da)}\right\rangle$ for each $g\in S^{1}(F)$. Fix an arbitrary $f_{0}\in S^{1}(F)$ for later.
\par
Let $(C_{n})_{n=1}^{\infty}\subseteq\mathscr{A}$ so $A\setminus C_{n}=\left\{\Xi\leq\widetilde{c}_{n}\right\}$ for some decreasing sequence $(\widetilde{c}_{n})_{n=1}^{\infty}\subseteq(-\infty,0)$ with $(\widetilde{c}_{n})_{n=1}^{\infty}\downarrow-\infty$ and
\begin{equation*}
    \left\vert{\int_{\left\{\Xi\leq\widetilde{c}_{n}\right\}}}\langle{x,f_{0}(a)}\rangle\mu(da)\right\vert\leq\frac{1}{n}
\end{equation*}
for all $n$. We may find a sequence of simple functions $(\Xi_{n})_{n=1}^{\infty}$ so $C_{n}\subseteq\{\Xi_{n}\leq\Xi\}$, $A\setminus C_{n}\subseteq\{\Xi_{n}\leq0\}$, and there exists a small $\delta>0$ so $\int_{A}\Xi_{n}(a)\mu(da)>\beta+\delta$ for sufficiently large $n$ (say, $n\geq n_{1}$).
\par
For any $n\in\mathbb{N}$, define a correspondence $G_{n}:A\longrightarrow 2^{E^{\ast}}$ by setting
\begin{equation*}
    G_{n}(a)=\left\{x^{\ast}:\langle{x,x^{\ast}}\rangle\geq\Xi_{n}(a)-\frac{1}{n}\right\}\cap F(a)
\end{equation*}
for $a\in C_{n}$, and $G_{n}(a)=\{f_{0}(a)\}$ for $a\in A\setminus C_{n}$. It is easy to see that $G_{n}$ is $\mathscr{A}$-measurable: if $U\subseteq E^{\ast}$ is $\sigma(E^{\ast},E)$-closed, then
\begin{equation*}
    G^{-1}_{n}(U)=\left(\left(\bigcup_{\alpha\in\Xi_{n}(A)}F^{-1}\left(U\cap\left\{x^{\ast}:\langle{x,x^{\ast}}\rangle\geq\alpha-\frac{1}{n}\right\}\right)\cap\left\{\Xi_{n}=\alpha\right\}\right)\cap C_{n}\right)
\end{equation*}
\begin{equation*}
    \cup \left(f_{0}^{-1}(U)\cap(A\setminus C_{n})\right)
\end{equation*}
which is evidently $\mathscr{A}$-measurable.
\par
Fix an arbitrary $n\in\mathbb{N}$. By the same argument as in the proof of Lemma \ref{lem:count-represent}, there exists an $\mathscr{A}$-measurable $h_{n}:A\longrightarrow E^{\ast}$ such that $h_{n}(a)\in G_{n}(a)$ for each $a\in A$. For $m\in\mathbb{N}$, define $B^{n,m}=h_{n}^{-1}(mB^{\ast})\in\mathscr{A}$, and set $h_{n,m}=\mathbf{1}_{B^{n,m}}h_{n}+\mathbf{1}_{A\setminus B^{n,m}}f_{0}$. Clearly, $h_{n,m}\in S^{1}(F)$. Notice that
\begin{equation*}
    \left\langle{x,\int_{A}h_{n,m}(a)\mu(da)}\right\rangle=\left\langle{x,\int_{B^{n,m}}h_{n}(a)\mu(da)}\right\rangle+\left\langle{x,\int_{A\setminus B^{n,m}}f_{0}(a)\mu(da)}\right\rangle
\end{equation*}
\begin{equation*}
    =\left\langle{x,\int_{B^{n,m}\cap C_{n}}}h_{n}(a)\mu(da)\right\rangle+\left\langle{x,\int_{B^{n,m}\cap(A\setminus C_{n})}}h_{n}(a)\mu(da)\right\rangle+\left\langle{x,\int_{A\setminus B^{n,m}}f_{0}(a)\mu(da)}\right\rangle
\end{equation*}
\begin{equation*}
    \geq\int_{B^{n,m}\cap C_{n}}\left(\Xi_{n}(a)-\frac{1}{n}\right)\mu(da)+\left\langle{x,\int_{\left(B^{n,m}\cap(A\setminus C_{n})\right)\cup\left(A\setminus B^{n,m}\right)}f_{0}(a)\mu(da)}\right\rangle.
\end{equation*}
Taking $m\to\infty$ in the last expression, we get $\int_{C_{n}}\left(\Xi_{n}(a)-\frac{1}{n}\right)\mu(da)+\int_{A\setminus C_{n}}\langle{x,f_{0}(a)}\rangle\mu(da)$, and from this limiting expression we obtain the chain of inequalities
\begin{equation*}
    \int_{C_{n}}\left(\Xi_{n}(a)-\frac{1}{n}\right)\mu(da)+\int_{A\setminus C_{n}}\langle{x,f_{0}(a)}\rangle\mu(da)\geq\int_{C_{n}}\left(\Xi_{n}(a)-\frac{1}{n}\right)\mu(da)-\frac{\mu(A\setminus C_{n})}{n}
\end{equation*}
\begin{equation*}
    =\int_{C_{n}}\Xi_{n}(a)\mu(da)-\frac{\mu(A)}{n}\geq\int_{A}\Xi_{n}(a)\mu(da)-\frac{\mu(A)}{n}.
\end{equation*}
So, for any $\varepsilon>0$, there exists $f_{n}\in S^{1}(F)$ (more precisely, we can take $f_{n}=h_{n,m}$ for sufficiently large $m$) so that
\begin{equation*}
    \left\langle{x,\int_{A}f_{n}(a)\mu(da)}\right\rangle+\varepsilon\geq\int_{A}\Xi_{n}(a)\mu(da)-\frac{\mu(A)}{n}.
\end{equation*}
Taking $\varepsilon=\frac{1}{n}$, it follows that there exists $f_{n}\in S^{1}(F)$ so that
\begin{equation}\label{eq:fn-sat}
    \left\langle{x,\int_{A}f_{n}(a)\mu(da)}\right\rangle\geq\int_{A}\Xi_{n}(a)\mu(da)-\frac{\mu(A)+1}{n}.
\end{equation}
Using the arbitrariness of $n\in\mathbb{N}$, we have a sequence $(f_{n})_{n=1}^{\infty}\subseteq S^{1}(F)$ so that each $f_{n}$ satisfies (\ref{eq:fn-sat}).
\par
Let $n\geq n_{1}$ be sufficiently large so that $\delta-\frac{\mu(A)+1}{n}>0$. Then
\begin{equation*}
    \left\langle{x,\int_{A}f_{n}(a)\mu(da)}\right\rangle\geq\int_{A}\Xi_{n}(a)\mu(da)-\frac{\mu(A)+1}{n}>\beta+\delta-\frac{\mu(A)+1}{n}>\beta.
\end{equation*}
Hence, there exists $f\in S^{1}(F)$ (namely, $f=f_{n}$ for some $n\geq n_{1}$, as above) so that $\left\langle{x,\int_{A}f_{n}(a)\mu(da)}\right\rangle>\beta$, a contradiction, as $\beta$ is assumed to satisfy the reverse inequality.
\end{proof}
\section{Proof of Theorem \ref{thm:dual-rep-represent}}\label{sec:prf-of-thm-rep}
In this section, we prove Theorem \ref{thm:dual-rep-represent}. Various tools are employed, including results from Appendix \ref{sec:tech}, and some results about acceptance sets (see \S\ref{subsec:accept} below).
\subsection{Acceptance Sets}\label{subsec:accept}
Given a risk measure $\varrho$, the acceptance set $\mathfrak{A}(\varrho)$ is defined by
\begin{equation*}
    \mathfrak{A}(\varrho)=\{\mathcal{X}:\varrho(\mathcal{X})\leq0\}.
\end{equation*}
The Fatou property implies that $\mathfrak{A}(\varrho)$ is $\sigma(L^{\infty},L^{1})$-closed (see \cite[Corollary 1]{melnikov-duality}).
\subsubsection{Characterizing the Acceptance Set}\label{subsubsec:char-accept}
In this subsection, we provide a characterization of the acceptance set of the integral infimal convolution of $(\varrho_{a})_{a\in A}$, in terms of the closure of a certain Aumann integral.
\par
Recall the Aumann integral from Definition \ref{def:aumann-int} in Appendix \ref{sec:tech}, which we reproduce here in a slightly less abstract setting. Given a correspondence $F:A\longrightarrow 2^{L^{\infty}(\mathbb{P})}$, an integrable selector of $F$ is an $\mathscr{A}$-measurable Gelfand integrable function $(X_{a})_{a\in A}\in\left(L^{\infty}(\mathbb{P})\right)^{A}$ such that $X_{a}\in F(a)$ for $\mu$-a.e. $a\in A$. The set of all integrable selectors of $F$ is denoted $S^{1}(F)$. The Aumann integral $\int_{A}F(a)\mu(da)$ of $F$ is defined as
\begin{equation*}
    \int_{A}F(a)\mu(da)=\left\{\int_{A}X_{a}\mu(da):(X_{a})_{a\in A}\in S^{1}(F)\right\}.
\end{equation*}
\begin{theorem}\label{thm:accept-set-characterize}
    Suppose $\operatorname{\Box}_{a\in A}\varrho_{a}\mu(da)$ is globally finite. Then, the acceptance set $\mathfrak{A}\left(\operatorname{\Box}_{a\in A}\varrho_{a}\mu(da)\right)$ of $\operatorname{\Box}_{a\in A}\varrho_{a}\mu(da)$ is the $L^{\infty}(\mathbb{P})$-closure of the Aumann integral $\int_{A}\mathfrak{A}(\varrho_{a})\mu(da)$.
\end{theorem}
\begin{proof}
    We follow the same idea as the proof of \cite[Theorem 4.1]{felixfs}, replacing finite Minkowski sums with Aumann integrals.
    \par
    Clearly, $\int_{A}\mathfrak{A}(\varrho_{a})\mu(da)\subseteq\mathfrak{A}\left(\operatorname{\Box}_{a\in A}\varrho_{a}\mu(da)\right)$. Since the integral infimal convolution is monotone and cash additive, we may apply the same argument as \cite[Lemma 4.3]{stoch-fin} to obtain that $\mathfrak{A}\left(\operatorname{\Box}_{a\in A}\varrho_{a}\mu(da)\right)$ is $L^{\infty}(\mathbb{P})$-closed, implying $\overline{\int_{A}\mathfrak{A}(\varrho_{a})\mu(da)}^{L^{\infty}}\subseteq\mathfrak{A}\left(\operatorname{\Box}_{a\in A}\varrho_{a}\mu(da)\right)$. Thus, it suffices to show the reverse inclusion.
    \par
    Let $\mathcal{X}\in\mathfrak{A}\left(\operatorname{\Box}_{a\in A}\varrho_{a}\mu(da)\right)$; denote $w=\left(\operatorname{\Box}_{a\in A}\varrho_{a}\mu(da)\right)(\mathcal{X})\leq0$. By cash additivity, $\left(\operatorname{\Box}_{a\in A}\varrho_{a}\mu(da)\right)\left(\mathcal{X}-w\right)=0$. Thus, there exists a sequence $\left((X^{n}_{a})_{a\in A}\right)_{n=1}^{\infty}\subseteq\mathbb{A}(\mathcal{X}-w)$ such that
    \begin{equation*}
        \lim_{n\to\infty}\int_{A}\varrho_{a}\left(X^{n}_{a}\right)\mu(da)=0.
    \end{equation*}
    Let $Y^{n}_{a}=X^{n}_{a}-\varrho_{a}(X^{n}_{a})$. Clearly, $(Y^{n}_{a})_{a\in A}\in\mathfrak{A}(\varrho_{a})$, implying that $\int_{A}Y^{n}_{a}\mu(da)\in\int_{A}\mathfrak{A}(\varrho_{a})\mu(da)$. Thus, since the $L^{\infty}(\mathbb{P})$-limit of $\left(\int_{A}Y^{n}_{a}\mu(da)\right)_{n=1}^{\infty}$ is $\mathcal{X}-w$, it follows that $\mathcal{X}-w\in\overline{\int_{A}\mathfrak{A}(\varrho_{a})\mu(da)}^{L^{\infty}}$. Since $w\leq0$, it follows that $\mathcal{X}\in\overline{\int_{A}\mathfrak{A}(\varrho_{a})\mu(da)}^{L^{\infty}}$, as desired.
\end{proof}
\subsubsection{Representing the Dual via Acceptance Sets}
In this subsection, we state a known result connecting acceptance sets to convex conjugates. This allows us to apply our results on the correspondence $a\longmapsto\mathfrak{A}(\varrho_{a})$ to dual representations.
\begin{lemma}\label{lem:dual-accept-set}
    Let $\varrho$ be any risk measure. Then, for any $\mathbb{Q}\in\mathscr{M}_{\mathbb{P}}$,
    \begin{equation*}
        \varrho^{\ast}(\mathbb{Q})=\sup_{\mathcal{X}\in\mathfrak{A}(\varrho)}\mathbb{E}^{\mathbb{Q}}(\mathcal{X}).
    \end{equation*}
\end{lemma}
\begin{proof}
    Fix $\mathbb{Q}\in\mathscr{M}_{\mathbb{P}}$; clearly, $\sup_{\mathcal{X}\in\mathfrak{A}(\varrho)}\mathbb{E}^{\mathbb{Q}}(\mathcal{X})\leq\varrho^{\ast}(\mathbb{Q})$. By cash additivity,
    \begin{equation*}
        \sup_{\mathcal{X}\in\mathfrak{A}(\varrho)}\mathbb{E}^{\mathbb{Q}}(\mathcal{X})\leq\varrho^{\ast}(\mathbb{Q})=\sup_{\mathcal{X}\in L^{\infty}(\mathbb{P})}\left(\mathbb{E}^{\mathbb{Q}}(\mathcal{X})-\varrho(\mathcal{X})\right)
    \end{equation*}
    \begin{equation*}
        =\sup_{\mathcal{X}\in L^{\infty}(\mathbb{P})}\left(\mathbb{E}^{\mathbb{Q}}(\mathcal{X}-\varrho(X))-\varrho(\mathcal{X}-\varrho(X))\right)=\sup_{\mathcal{X}\in\{\varrho=0\}}\mathbb{E}^{\mathbb{Q}}\left(\mathcal{X}\right)\leq\sup_{\mathcal{X}\in\mathfrak{A}(\varrho)}\mathbb{E}^{\mathbb{Q}}(\mathcal{X})
    \end{equation*}
    implying that $\sup_{\mathcal{X}\in\mathfrak{A}(\varrho)}\mathbb{E}^{\mathbb{Q}}(\mathcal{X})\leq\varrho^{\ast}(\mathbb{Q})\leq\sup_{\mathcal{X}\in\mathfrak{A}(\varrho)}\mathbb{E}^{\mathbb{Q}}(\mathcal{X})$, showing that $\varrho^{\ast}(\mathbb{Q})=\sup_{\mathcal{X}\in\mathfrak{A}(\varrho)}\mathbb{E}^{\mathbb{Q}}(\mathcal{X})$. Since $\mathbb{Q}\in\mathscr{M}_{\mathbb{P}}$ was arbitrary, this proves the claim.
\end{proof}
\subsubsection{Measurability of the Acceptance Set Correspondence}
In this subsection, we establish that the acceptance set correspondence $a\longmapsto\mathfrak{A}(\varrho_{a})$ is $\mathscr{A}$-measurable in the sense of Definition \ref{def:corr-measurable} from Appendix \ref{sec:tech}. We assume in this subsection that $(\varrho_{a})_{a\in A}$ denotes an $\mathscr{A}$-measurable indexed collection of risk measures such that each $\varrho_{a}$ has the Lebesgue property.
\begin{lemma}\label{lem:measurable-corr-accept}
    Let $U\subseteq L^{\infty}(\mathbb{P})$ be $\sigma(L^{\infty},L^{1})$-closed. Then
    \begin{equation*}
        \left\{a\in A:U\cap\mathfrak{A}(\varrho_{a})\neq\emptyset\right\}\in\mathscr{A}.
    \end{equation*}
\end{lemma}
\begin{proof}
    It is no loss of generality to assume that $U$ is bounded in $L^{\infty}(\mathbb{P})$, since one can write $U$ as a countable union of closed and $L^{\infty}(\mathbb{P})$-bounded sets. Furthermore, we may assume that $U\neq\emptyset$ (if $U=\emptyset$, the claim would be trivial).
    \par
    We claim that
    \begin{equation}\label{eq:sets-do-be-equal}
        \left\{a\in A:U\cap\mathfrak{A}(\varrho_{a})\neq\emptyset\right\}=\left\{a\in A:\inf_{\mathcal{Y}\in U}\varrho_{a}(\mathcal{Y})\leq0\right\}.
    \end{equation}
    Clearly, $\left\{a\in A:U\cap\mathfrak{A}(\varrho_{a})\neq\emptyset\right\}\subseteq\left\{a\in A:\inf_{\mathcal{Y}\in U}\varrho_{a}(\mathcal{Y})\leq0\right\}$. Thus, it suffices to show the reverse inclusion. If $\inf_{\mathcal{Y}\in U}\varrho_{a}(\mathcal{Y})\leq0$, there exists $(\mathcal{Y}^{n})_{n=1}^{\infty}\subseteq U$ such that
    \begin{equation*}
        \varrho_{a}\left(\mathcal{Y}^{n}\right)\leq\frac{1}{n}.
    \end{equation*}
    Using the Banach-Alaoglu theorem, $L^{\infty}(\mathbb{P})$-boundedness of $U$, and $\sigma(L^{\infty},L^{1})$-closedness of $U$, we may find a subsequence $(n_{k})_{k=1}^{\infty}$ such that $(\mathcal{Y}^{n_{k}})_{k=1}^{\infty}$ converges to some $\mathcal{Z}\in U$ in $\sigma(L^{\infty},L^{1})$. For each $n$, there exists $k_{0}$ such that $k\geq k_{0}$ implies $\mathcal{Y}^{n_{k}}\in\{\varrho_{a}\leq\frac{1}{n}\}$. The Fatou property implies the set $\left\{\varrho_{a}\leq\frac{1}{n}\right\}$ is $\sigma(L^{\infty},L^{1})$-closed, and we therefore have that $\mathcal{Z}\in\left\{\varrho_{a}\leq\frac{1}{n}\right\}$ for each $n$. Thus,
    \begin{equation*}
        \mathcal{Z}\in\bigcap_{n\in\mathbb{N}}\left\{\varrho_{a}\leq\frac{1}{n}\right\}=\{\varrho_{a}\leq0\}.
    \end{equation*}
    By the above argument, there exists $\mathcal{Z}\in U$ with $\varrho_{a}(\mathcal{Z})\leq0$, implying that $a\in\left\{b\in A:U\cap\mathfrak{A}(\varrho_{b})\neq\emptyset\right\}$, as desired.
    \par
    As a consequence of (\ref{eq:sets-do-be-equal}), it suffices to show that $a\longmapsto\inf_{\mathcal{Y}\in U}\varrho_{a}(\mathcal{Y})$ is $\mathscr{A}$-measurable. Let $V\subseteq U$ be a countable dense set for the topology $\tau_{L^{0}}$ of convergence in probability restricted to $U$ (such a set exists, since $(\Omega,\mathscr{F},\mathbb{P})$ is separable). We claim that
    \begin{equation}\label{eq:infs-equal-dense}
        \inf_{\mathcal{Y}\in U}\varrho_{a}(\mathcal{Y})=\inf_{\mathcal{Y}\in V}\varrho_{a}(\mathcal{Y}),
    \end{equation}
    for all $a\in A$, which would prove the claim, since $a\longmapsto\inf_{\mathcal{Y}\in U}\varrho_{a}(\mathcal{Y})$ would be a countable infimum of $\mathscr{A}$-measurable functions. Since $\varrho_{a}$ has the Lebesgue property, $U$ is $L^{\infty}(\mathbb{P})$-bounded, and $\overline{V}^{\tau_{L^{0}}}=U$, (\ref{eq:infs-equal-dense}) holds.
\end{proof}
\subsection{The Proof}
\begin{proof}[Proof of Theorem \ref{thm:dual-rep-represent}]
    Recall that
    \begin{equation*}
        \left(\operatorname{\Box}_{a\in A}\varrho_{a}\mu(da)\right)^{\ast}(\mathbb{Q})=\sup_{\mathcal{X}\in L^{\infty}(\mathbb{P})}\left(\mathbb{E}^{\mathbb{Q}}(\mathcal{X})-\left(\operatorname{\Box}_{a\in A}\varrho_{a}\mu(da)\right)(\mathcal{X})\right)
    \end{equation*}
    for each $\mathbb{Q}\in \mathscr{M}_{\mathbb{P}}$. By Lemma \ref{lem:dual-accept-set} and Theorem \ref{thm:accept-set-characterize},
    \begin{equation}\label{eq:phi-accept-int}
        \left(\operatorname{\Box}_{a\in A}\varrho_{a}\mu(da)\right)^{\ast}(\mathbb{Q})=\sup_{\mathcal{X}\in\overline{\int_{A}\mathfrak{A}(\varrho_{a})\mu(da)}^{L^{\infty}}}\mathbb{E}^{\mathbb{Q}}\left(\mathcal{X}\right).
    \end{equation}
    We claim that $S^{1}(F)\neq\emptyset$ ($S^{1}(F)$ is defined in \S\ref{subsubsec:char-accept}), where $F$ is the correspondence $F=\left(a\longmapsto\mathfrak{A}(\varrho_{a})\right)$. Define $(X_{a})_{a\in A}\in\left(L^{\infty}(\mathbb{P})\right)^{A}$ by setting
    \begin{equation*}
        X_{a}=-\varrho_{a}(0).
    \end{equation*}
    It is easy to see that $(X_{a})_{a\in A}$ is $\mathscr{A}$-measurable; furthermore, since $\int_{A}\vert{\varrho_{a}(0)}\vert\mu(da)<\infty$, $(X_{a})_{a\in A}$ is Gelfand integrable. For each $a\in A$, we have that
    \begin{equation*}
        \varrho_{a}(X_{a})=\varrho_{a}(-\varrho_{a}(0))=\varrho_{a}(0)+(-\varrho_{a}(0))=0\leq0
    \end{equation*}
    by cash additivity. Thus, $X_{a}\in F(a)$ for each $a\in A$. These facts together imply that $(X_{a})_{a\in A}\in S^{1}(F)$, showing that $S^{1}(F)\neq\emptyset$.
    \par
    By Lemma \ref{lem:measurable-corr-accept}, $F$ is $\mathscr{A}$-measurable in the sense of Definition \ref{def:corr-measurable} from Appendix \ref{sec:tech}. Thus, since $S^{1}(F)\neq\emptyset$, the preconditions for Theorem \ref{thm:hiai-umegaki} and Theorem \ref{thm:hiai-umegaki-dense} are met. Noting that the norm topology on $L^{\infty}(\mathbb{P})$ is finer than $\sigma(L^{\infty},L^{1})$, Theorem \ref{thm:hiai-umegaki-dense} and (\ref{eq:phi-accept-int}) imply that
    \begin{equation*}
        \left(\operatorname{\Box}_{a\in A}\varrho_{a}\mu(da)\right)^{\ast}(\mathbb{Q})=\int_{A}\sup_{\mathcal{X}\in F(a)}\mathbb{E}^{\mathbb{Q}}\left(\mathcal{X}\right)\mu(da),
    \end{equation*}
    where we note that $a\longmapsto\sup_{\mathcal{X}\in F(a)}\mathbb{E}^{\mathbb{Q}}\left(\mathcal{X}\right)$ is $\mathscr{A}$-measurable, by virtue of Lemma \ref{lem:sup-measurable}. By Lemma \ref{lem:dual-accept-set}, we have that $\varrho^{\ast}_{a}(\mathbb{Q})=\sup_{\mathcal{X}\in F(a)}\mathbb{E}^{\mathbb{Q}}\left(\mathcal{X}\right)$ for each $a\in A$, which implies that $a\longmapsto\varrho^{\ast}(\mathbb{Q})$ is $\mathscr{A}$-measurable, and that
    \begin{equation*}
        \left(\operatorname{\Box}_{a\in A}\varrho_{a}\mu(da)\right)^{\ast}(\mathbb{Q})=\int_{A}\varrho^{\ast}_{a}(\mathbb{Q})\mu(da),
    \end{equation*}
    as desired.
\end{proof}
\section{Proof of Theorem \ref{thm:attain-notattain}}\label{sec:prf-attain-notattain}
In this section, we prove the second part of Theorem \ref{thm:attain-notattain}, showing that under certain circumstances the risk sharing problem is not well-posed.
\subsection{Lemmata}
Proposition \ref{prop:inflate-continuous} establishes concavity of inflated risk measures in the inflation parameter $\gamma$. In this subsection we recount a particular consequence of concavity—that constancy on a subinterval implies constancy elsewhere. We begin with a general formulation of the principle.
\begin{lemma}\label{lem:concave-function-constant}
    Let $\varphi:I\longrightarrow \mathbb{R}$ be a non-decreasing concave function on an interval $I\subseteq\mathbb{R}$. If $\gamma,\gamma',\gamma''\in I$, $\gamma''\geq\gamma'>\gamma$, and $\varphi(\gamma)=\varphi(\gamma')$, then $\varphi(\gamma)=\varphi(\gamma'')$.
\end{lemma}
\begin{proof}
    The case $\gamma''=\gamma'$ is trivial, so we assume $\gamma''>\gamma'$. The secant slope inequality for concave functions yields
    \begin{equation*}
        0=\frac{\varphi(\gamma')-\varphi(\gamma)}{\gamma'-\gamma}\geq\frac{\varphi(\gamma'')-\varphi(\gamma')}{\gamma''-\gamma'}\geq0,
    \end{equation*}
    where the last inequality follows from $\varphi$ being non-decreasing. Hence
    \begin{equation*}
        \frac{\varphi(\gamma'')-\varphi(\gamma')}{\gamma''-\gamma'}=0,
    \end{equation*}
    implying $\varphi(\gamma)=\varphi(\gamma')=\varphi(\gamma'')$, as desired.
\end{proof}
Applying Lemma \ref{lem:concave-function-constant} to the $\gamma$-inflation of a coherent risk measure with the Lebesgue property yields the following specialization.
\begin{lemma}\label{lem:constant-inflation}
    Let $\varrho$ be a risk measure with the Lebesgue property and the representation (\ref{eq:dualsetrep}), and fix a random variable $\mathcal{X}\in L^{\infty}(\mathbb{P})$. If $\gamma,\gamma',\gamma''\in[1,\infty)$, $\gamma''\geq\gamma'>\gamma$, and $\widetilde{\varrho}_{\gamma}(\mathcal{X})=\widetilde{\varrho}_{\gamma'}(\mathcal{X})$, then $\widetilde{\varrho}_{\gamma''}(\mathcal{X})=\widetilde{\varrho}_{\gamma}(\mathcal{X})$.
\end{lemma}
\begin{proof}
    Define $\varphi(\gamma)=\widetilde{\varrho}_{\gamma}(\mathcal{X})$ for $\gamma\in[1,\infty)$. Proposition \ref{prop:inflate-continuous} implies $\varphi$ is concave and non-decreasing, and the claim now follows immediately from Lemma \ref{lem:concave-function-constant}.
\end{proof}
We now state an interesting companion result to Lemma \ref{lem:constant-inflation}, which implies a random variable whose law is non-atomic always satisfies the second precondition for non-existence in Theorem \ref{thm:attain-notattain}.
\begin{proposition}\label{prop:essup-and-inflation}
    Suppose $\mathcal{X}\in L^{\infty}(\mathbb{P})$ satisfies $\mathbb{P}\left(\left\{\mathcal{X}=\overline{s}\right\}\right)=0$, where $\overline{s}=\mathbb{P}-\mathrm{essup}\mathcal{X}$, and $\varrho$ is a risk measure with the Lebesgue property and the representation (\ref{eq:dualsetrep}). Then the map $\gamma\longmapsto\widetilde{\varrho}_{\gamma}(\mathcal{X})$ is strictly increasing on $[1,\infty)$. In particular, for every $\gamma\in[1,\infty)$, there exists $\gamma'>\gamma$ such that $\widetilde{\varrho}_{\gamma}(\mathcal{X})\neq\widetilde{\varrho}_{\gamma'}(\mathcal{X})$.
\end{proposition}
\begin{proof}
    Supposing for contradiction the failure of strict monotonicity, there would exist $\gamma,\gamma'\in[1,\infty)$ with $\gamma<\gamma'$ and $\widetilde{\varrho}_{\gamma}(\mathcal{X})=\widetilde{\varrho}_{\gamma'}(\mathcal{X})$. Lemma \ref{lem:constant-inflation} conjoined with Proposition \ref{prop:inflate-continuous} imply $\widetilde{\varrho}_{\gamma}(\mathcal{X})=\overline{s}$.
    \par
    As noted in the proof of Theorem \ref{thm:inflate-dual}, every inflation of $\varrho$ must have the Lebesgue property. So the Jouini-Schachermayer-Touzi theorem \cite[Theorem 2.4]{owari-jfa} implies the existence of $\mathbb{Q}\in\mathscr{M}_{\mathbb{P}}$ with
    \begin{equation}\label{eq:essup-is-expectation}
        \overline{s}=\widetilde{\varrho}_{\gamma}(\mathcal{X})=\mathbb{E}^{\mathbb{Q}}(\mathcal{X}).
    \end{equation}
    Since $\mathbb{Q}\ll\mathbb{P}$ and $\mathbb{P}\left(\left\{\mathcal{X}=\overline{s}\right\}\right)=0$, $\mathbb{Q}\left(\left\{\mathcal{X}=\overline{s}\right\}\right)=0$. Hence there exists $\varepsilon>0$ such that, denoting $\delta_{\varepsilon}=\mathbb{Q}\left(\left\{\mathcal{X}\leq\overline{s}-\varepsilon\right\}\right)$, we have that $\delta_{\varepsilon}>0$. Evidently,
    \begin{equation*}
        \mathbb{E}^{\mathbb{Q}}(\mathcal{X})\leq\delta_{\varepsilon}\left(\overline{s}-\varepsilon\right)+(1-\delta_{\varepsilon})\overline{s}=\overline{s}-\delta_{\varepsilon}\varepsilon<\overline{s},
    \end{equation*}
    which contradicts (\ref{eq:essup-is-expectation}).
\end{proof}
\subsection{The Proof}
\begin{proof}[Proof of Theorem \ref{thm:attain-notattain}]
We prove the second assertion. Assume, for contradiction, that $(X_{a})_{a\in A}$ is optimal and the preconditions of the second assertion are satisfied.
\par
For each $n\in\mathbb{N}$, define
\begin{equation*}
    B_{n}=\left\{a\in A:\gamma_{a}\geq\Gamma+\frac{1}{n}\right\},
\end{equation*}
\begin{equation*}
    C_{n}=\left\{a\in A:\gamma_{a}<\Gamma+\frac{1}{n}\right\}.
\end{equation*}
As $(X_{a})_{a\in A}$ is optimal, $\widetilde{\varrho}_{\Gamma}(\mathcal{X})=\int_{A}\widetilde{\varrho}_{\gamma_{a}}(X_{a})\mu(da)$ by Theorem \ref{thm:inflate-dual}. Hence
\begin{equation*}
    \widetilde{\varrho}_{\Gamma}(\mathcal{X})=\int_{B_{n}}\widetilde{\varrho}_{\gamma_{a}}(X_{a})\mu(da)+\int_{C_{n}}\widetilde{\varrho}_{\gamma_{a}}(X_{a})\mu(da)
\end{equation*}
\begin{equation*}
    \geq\widetilde{\varrho}_{\Gamma}\left(\int_{C_{n}}X_{a}\mu(da)\right)+\widetilde{\varrho}_{\Gamma+\frac{1}{n}}\left(\int_{B_{n}}X_{a}\mu(da)\right).
\end{equation*}
Suppose, for some $n\in\mathbb{N}$, that $\widetilde{\varrho}_{\Gamma+\frac{1}{n}}\left(\int_{B_{n}}X_{a}\mu(da)\right)>\widetilde{\varrho}_{\Gamma}\left(\int_{B_{n}}X_{a}\mu(da)\right)$. Then
\begin{equation*}
    \widetilde{\varrho}_{\Gamma}(\mathcal{X})\geq\widetilde{\varrho}_{\Gamma}\left(\int_{C_{n}}X_{a}\mu(da)\right)+\widetilde{\varrho}_{\Gamma+\frac{1}{n}}\left(\int_{B_{n}}X_{a}\mu(da)\right)
\end{equation*}
\begin{equation*}
    >\widetilde{\varrho}_{\Gamma}\left(\int_{C_{n}}X_{a}\mu(da)\right)+\widetilde{\varrho}_{\Gamma}\left(\int_{B_{n}}X_{a}\mu(da)\right)\geq\widetilde{\varrho}_{\Gamma}(\mathcal{X})
\end{equation*}
by the subadditivity of coherent risk measures, a contradiction. So, for all $n\in\mathbb{N}$, $\widetilde{\varrho}_{\Gamma+\frac{1}{n}}\left(\int_{B_{n}}X_{a}\mu(da)\right)=\widetilde{\varrho}_{\Gamma}\left(\int_{B_{n}}X_{a}\mu(da)\right)$. Lemma \ref{lem:constant-inflation} implies
\begin{equation}\label{eq:inflation-equality}
    \widetilde{\varrho}_{\Gamma}\left(\int_{B_{n}}X_{a}\mu(da)\right)=\widetilde{\varrho}_{\Gamma'}\left(\int_{B_{n}}X_{a}\mu(da)\right)
\end{equation}
for all sufficiently large $n\in\mathbb{N}$ (large enough so that $\Gamma'\geq\Gamma+\frac{1}{n}$).
\par
Define a linear operator $G:L^{1}(\mathbb{P})\longrightarrow L^{1}(\mu)$ by $G(\mathcal{Y})=\mathbb{E}^{\mathbb{P}}(X_{a}\mathcal{Y})$ for each $\mathcal{Y}\in L^{1}(\mathbb{P})$. In view of the closed graph theorem, $G$ is norm to norm continuous, and hence weak to weak continuous (see \cite[Theorem 3.10]{brezis}). Hence, if $D$ denotes the unit ball of $L^{\infty}(\mathbb{P})$, the Dunford-Pettis theorem implies $G(D)$ is uniformly $\mu$-integrable. So, for any $\varepsilon>0$, there exists $\delta>0$ so $\mu(E)<\delta$ implies
\begin{equation*}
    \left\Vert{\int_{E}X_{a}\mu(da)}\right\Vert_{L^{1}(\mathbb{P})}=\sup_{\mathcal{Y}\in D}\left\vert\int_{E}\mathbb{E}^{\mathbb{P}}(X_{a}\mathcal{Y})\mu(da)\right\vert<\varepsilon.
\end{equation*}
As $n\to\infty$, $\mu(C_{n})\to0$, and in particular $\mu(C_{n})<\delta$ for large enough $n$, so that
\begin{equation*}
    \left\Vert{\mathcal{X}-\int_{B_{n}}X_{a}\mu(da)}\right\Vert_{L^{1}(\mathbb{P})}=\left\Vert{\int_{C_{n}}X_{a}\mu(da)}\right\Vert_{L^{1}(\mathbb{P})}<\varepsilon
\end{equation*}
for large enough $n$. As $\varepsilon>0$ was arbitrary, we have that $\left(\int_{B_{n}}X_{a}\mu(da)\right)_{n=1}^{\infty}$ converges to $\mathcal{X}$ in $L^{1}(\mathbb{P})$, and hence also in probability. Further, $\left(\int_{B_{n}}X_{a}\mu(da)\right)_{n=1}^{\infty}$ evidently converges in $\sigma(L^{\infty},L^{1})$ to $\mathcal{X}$, so the Banach-Steinhaus theorem implies $\left(\int_{B_{n}}X_{a}\mu(da)\right)_{n=1}^{\infty}$ is bounded in $L^{\infty}(\mathbb{P})$. We can thus apply the Lebesgue property of $\widetilde{\varrho}_{\Gamma}$ and $\widetilde{\varrho}_{\Gamma'}$ to obtain
\begin{equation*}
    \widetilde{\varrho}_{\Gamma}\left(\mathcal{X}\right)=\widetilde{\varrho}_{\Gamma'}\left(\mathcal{X}\right)
\end{equation*}
by taking $n\to\infty$ in (\ref{eq:inflation-equality}), contradicting the assumption $\widetilde{\varrho}_{\Gamma}\left(\mathcal{X}\right)\neq\widetilde{\varrho}_{\Gamma'}\left(\mathcal{X}\right)$.
\end{proof}
\section*{Acknowledgments}
The author thanks two anonymous reviewers and an editor for their pertinent feedback and comments.
\printbibliography
\end{document}